\documentclass[proof]{pasj00}
\usepackage{natbib}
\draft

\begin{document}
\SetRunningHead{S. Takasao et al.}{Acceleration Mechanisms of Chromospheric Jets Associated with Emerging Flux}

\title{Numerical Simulations of Solar Chromospheric Jets Associated with Emerging Flux}

\author{
Shinsuke \textsc{Takasao}\altaffilmark{1,2},
Hiroaki  \textsc{Isobe}\altaffilmark{3}, and 
Kazunari  \textsc{Shibata}\altaffilmark{1}
}
\email{takasao@kwasan.kyoto-u.ac.jp}
\altaffiltext{1}{Kwasan and Hida Observatories, Kyoto University, 
Yamashina, Kyoto 607-8471, Japan}
\altaffiltext{2}{Department of Astronomy, Kyoto University, 
Sakyo, Kyoto 606-8502, Japan}
\altaffiltext{3}{Unit of Synergetic Studies for Space, Kyoto University,
Yamashina, Kyoto 607-8471, Japan}

%


\maketitle

\begin{abstract}
We studied the acceleration mechanisms of chromospheric jets associated
with emerging flux using a two dimensional magnetohydrodynamic (MHD) simulation.
We found that slow mode shock waves generated by magnetic reconnection
in the chromosphere and the photosphere play key roles in the acceleration
mechanisms of chromospheric jets.
An important parameter is the height of magnetic reconnection.
When magnetic reconnection takes place near the photosphere,
the reconnection outflow collides with the region where the plasma beta is much larger than unity.
Then the plasma moves along a magnetic field.
This motion generates a slow mode wave.
The slow mode wave develops to a strong slow shock as it propagates upward.
When the slow shock crosses the transition region, the transition region is lifted up.
As a result, we obtain a chromospheric jet as the lifted transition region.
When magnetic reconnection takes place in the upper chromosphere,
the chromospheric plasma is accelerated due to the combination of the Lorentz force and the whip-like motion
of magnetic field.
We found that the chromospheric plasma is further accelerated through the interaction between the transition region
(steep density gradient) and a slow shock emanating from the reconnection point.
This is an MHD effect which has not been discussed before.
\end{abstract}

\section{Introduction}
\begin{figure}
  \begin{center}
    \FigureFile(80mm,80mm){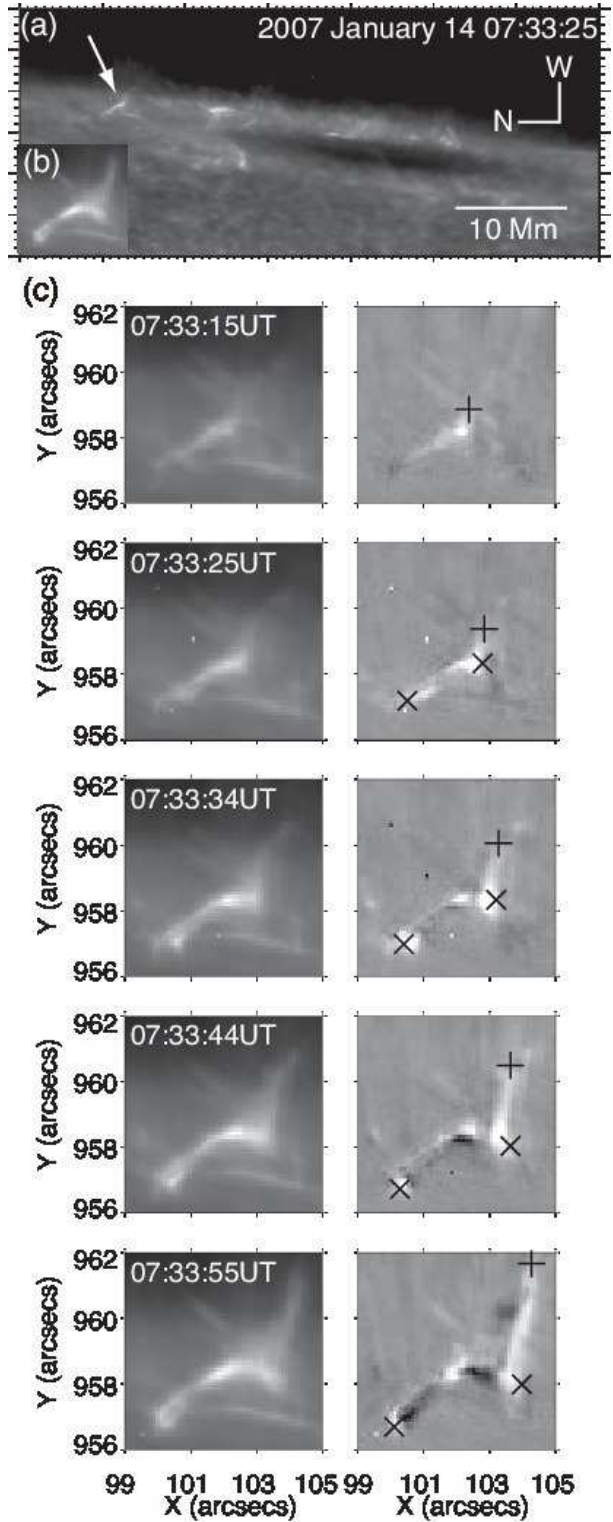}
  \end{center}
  \caption{(a) Ca \emissiontype{II} H broadband filter image of the active region NOAA 10935. (b) Enlarged image of the chromospheric anemone jet indicated by the white arrow in (a). (c) Left: Time-sequenced images of the chromospheric jet (Ca \emissiontype{II}) shown in (b). Right: Running difference images of the jet. Cross marks indicate the heads of the apparent horizontal motion along the loop structure and plus marks indicate the heads of the vertical jet. 1~arcsec$\sim$730~km.}\label{overview}
\end{figure}

Collimated ejections of chromospheric plasma into coronal heights have been studied for many years.
Spicules are jet-like structures seen in the limb of the Sun \citep{ste00, ana10}.
Surges are chromospheric jets seen dark in H$\alpha$ images and their heights reach up to $\sim200$~Mm  \citep{roy73,yos03}.
Recent observations have found the tiny chromospheric jets with the loop structures at their foot-points \citep{shi07}.
They are called  the chromospheric anemone jets.
Their typical height is $1- 4$~Mm \citep{nis11}.
Figure~\ref{overview} shows an example of the jets observed by the Solar Optical Telescope \citep{tsu08}  on board the {\it Hinode} satellite \citep{kos07}.

\par
Chromospheric jets are often observed in emerging flux regions and around
sunspots. They are cool ($\sim10^4$~K) compared to coronal plasma ($> 10^6$~K).
Many chromospheric jets disappear in the images of chromospheric lines before returning to the chromosphere. This may be due to some heating process or rapid expansion of the chromospheric material of jets (the rapid expansion is shown in this paper).
Therefore for correct interpretations of observations we should understand how chromospheric plasma is ejected.

\par
Trigger of chromospheric jets can lie at various height in the solar atmosphere.
X-ray observations found coronal hot jets (X-ray jets) associated with magnetic loops \citep{shi94}, which implies that the energy release process takes place in the corona.
Multi-wavelength coronal observations have shown that some cool jets (surges) coexist with hot jets \citep{sch95,can96,cha99}, which implies that the trigger is located in the lower corona or in the upper chromosphere.
It is known that some surges are associated with the Ellerman bombs or moustaches \citep{ell17, rust68}, where the Ellerman bombs are sudden energy releases in the low chromosphere or in the photosphere and their emissions are bright at both wings of the H$\alpha$ line \citep{kit83,fan06,wat08,mat08}. This implies that they are triggered in the low chromosphere or in the photosphere.

\par
Using MHD simulations,  \citet{yok96} studied the acceleration mechanism of surges. 
They simulated the evolution of an emerging flux in a uniform coronal magnetic field.
They obtained both hot and cool jets (which can correspond to the X-ray jets and surges, respectively) simultaneously as results of magnetic reconnection between the emerging flux and the pre-existing coronal field, where magnetic reconnection is a physical process in which a magnetic field in a highly conducting plasma changes its topology due to finite resistivity.
Their results are consistent with multi-wavelength observations.
They pointed out that the hot jets and the cool jets are accelerated mainly due to the gas pressure gradient of the heated plasma and the Lorentz force, respectively. 
Note that the chromospheric evaporation due to the heat conduction is also important for the acceleration of the X-ray jets \citep{smj00}.
More elaborated modeling for coronal jets has been performed by other authors \citep{iso06,pariat09,arc10,jia12}. 
\citet{mar11} proposed that strong Lorentz force due to large magnetic field gradient squeezes chromospheric plasma and produces spicule-like jets moving at a local sound speed.
\citet{mrn08} compared a three-dimensional numerical experiment with an observed coronal jet and showed that their numerical results are consistent with the observations.

\par
When magnetic loops expand in the solar atmosphere, magnetic reconnection can take place at various heights.
Now we should note that in the chromosphere the gravitational stratification is much stronger than in the corona.
As a result, the plasma beta varies from much larger than unity to less than unity. 
Therefore, the physical processes associated with magnetic reconnection can be different at various heights in the solar atmosphere.

\par
Can the lower atmospheric reconnection produce tall ($\sim$1-10~Mm) jets?
For example, observations have suggested that chromospheric anemone jets are considered as results of magnetic reconnection in the chromosphere \citep{mor10, sin11}.
Their speed is comparable to the expected local Alfv\'en speed ($\sim 10$~km~s$^{-1}$) in the upper chromosphere.
Suppose that magnetic reconnection takes place in the upper chromosphere.
Then it is highly possible that jets are reconnection outflows accelerated due to the whip-like motion of magnetic field, considering that the outflow speed is almost the same as the local Alfv\'en speed.
Next, suppose that magnetic reconnection occurred in the middle or low chromosphere where the plasma beta is expected to be larger than unity.
Then it is difficult to regard observed jets as reconnection outflows because the Alfv\'en velocity is lower than the observed velocity.

\par
Let us estimate the maximum height which reconnection outflows can reach.
Suppose that the kinetic energy is converted into the potential energy. Then we obtain $\rho V_{\it A}^2/2 \sim \rho g H_{jet}$, where $\rho$, $V_{\it A}$, $g$ and $H_{jet}$ are the density, the Alfv\'en velocity, the gravitational acceleration and the height of the jet, respectively. 
We obtain $H_{jet} \sim (p/\rho g) (B^2/(8\pi p)) \sim H_p/\beta$, where $p$, $B$, $H_p$ and $\beta$ are the gas pressure, the magnetic field strength, the pressure scale height and the plasma beta in the chromosphere, respectively.
Suppose, for example, $\beta \sim 1$. Then the height of the jet will be $H\sim 200$~km, which is much shorter than the typical height of observed jets (e.g. $1-4\times10^3$~km for the chromospheric anemone jets).
Therefore we need to consider another scenario in which only a fraction of plasma is accelerated.

\par
A promising model is the acceleration mechanism through the interaction between a slow shock and the transition region \citep{ost61,shi82,shi07,sue82,dep04,heg07}.
The transition region can be launched when a slow shock pass through it.
The ejected chromospheric plasma behind the contact discontinuous layer is regarded as chromospheric jets. 
\citet{shi07} pointed out that slow mode waves can be generated by magnetic reconnection between small emerging flux and the pre-existing ambient field. Then they proposed that the slow mode waves become shocks for some reason and produce chromospheric jets. 
Magnetic reconnection can generate waves of various kinds \citep{iso08,nis08,he09,kig10}.
The wave can become shocks through non-linear processes.
\citet{car97} compared observations with simulations and found that slow shocks are ubiquitously generated in the chromosphere.
As supported in the above studies, the slow shock acceleration mechanism has the potential to systematically account for observational characteristics of chromospheric jets.

\par
The slow shock acceleration mechanism has been studied by many authors.
\citet{sue82} and \citet{shi82} studied the mechanism using one-dimensional hydrodynamic simulations. 
Their results are both qualitatively and quantitatively consistent with observational properties of spicules and H$\alpha$ surges. 
They also proposed the possibility that the Ellerman bombs generate slow shocks and lead to surges.
\citet{sai01} showed that the Alfv\'en waves also produce slow shocks due to a non-linear coupling.
Although both slow and fast mode waves are generated through a non-linear process of the Alfv\'en waves, their results suggest that slow mode waves play more important roles in generation of jets than fast mode waves.
Some authors discussed the slow shock mechanism by including other effects \citep{ste93,heg07,mar09,mur11}.
\citet{han06} and \citet{dep07} discussed the relation between chromospheric jets and the slow mode shocks generated by convective flows and global p-mode oscillations.

\par
In the previous studies, waves are generated in the chromosphere or in the photosphere, artificially \citep[e.g.][]{shi82}, by using observational data of the photospheric motions \citep[e.g.][]{mat08}, and convective flows in a self-consistent manner \citep[e.g.][]{han06}.
Although \citet{mar09} suggested the possibility that sudden energy release due to reconnection in the low atmosphere can lead to chromospheric jets, they did not identify the connection.
The relation between the acceleration of chromospheric jets and magnetic reconnection has been still unclear.

\par
Even if magnetic reconnection takes place in the upper chromosphere, the interaction between slow shocks and the transition region can play an important role in accelerating chromospheric jets.
The standing slow shocks will emanate from a reconnection point if the resistivity is localized in some way (the Petschek-type reconnection; \citet{pet64}).
For a detailed discussion on the possibility of the localization of the resistivity, see Section~\ref{anom}.
Suppose that the Petschek-type reconnection occurred in the upper chromosphere. 
Then the reconnection inflow advects the transition region to the reconnection region, particularly to the slow shocks. 
As a result, the slow shock can cross the transition region.
The interaction can change the flow structure of the reconnection outflow.
Now we have less knowledge of the interaction processes in MHD shock cases.
Therefore the interaction makes it difficult for us to correctly understand the dynamics of jets.

\par
To self-consistently account for the acceleration mechanisms of chromospheric jets associated with emerging flux, we performed a two-dimensional (2D) MHD simulation.
In particular, we focused on the relation between magnetic reconnection and chromospheric jets.
Magnetic reconnection occurred between emerging magnetic loops and the pre-existing field in the chromosphere and at the foot-points of the emerging flux.
Slow shocks are generated by magnetic reconnection in the lower atmosphere (near the photosphere) and upper chromosphere.
The slow shocks accelerate the chromospheric plasma along a magnetic field, producing chromospheric jets.
The key process in this paper is the interaction between slow shocks and the transition region.
An important point we found is that the reconnection height determines the acceleration mechanism of chromospheric jets.
To investigate the interaction process in detail, we also performed numerous 1.5D simulations, as presented in Appendix.

\par
We note that evolution of emerging flux in a 3D space is similar to that in a 2D space.
For emerging process, compare \citet{shi89} (2D) and \citet{mat93} (3D) and for Ellerman bombs, \citet{iso07} (2D) and \citet{arc09} (3D).
In 3D we will find a wider variety of physics than in 2D, including the interchange instability \citep{iso06} and magnetic reconnection between not perfectly anti-paralleled field lines \citep{jia11,nak12}. 
Here we focus on fundamental physics in a 2D space.

\par
In Section \ref{sec:setup}, we introduce the numerical model adopted here.
In Section \ref{sec:result}, we present results of our 2D numerical simulation. 
Here we separately describe the acceleration processes of the two jets Jet~A and B 
to clarify that the height of magnetic reconnection determines the acceleration mechanism.
To investigate the acceleration process of Jet~B in detail, we performed numerous 1.5D simulations, which is summarized in Appendix.
In Section \ref{sec:discussion} we present a schematic diagram summarizing the acceleration mechanisms of chromospheric jets.
Here we also give implications for observations and discuss the validity of the adopted assumptions.

\section{Numerical Setup} \label{sec:setup}

\subsection{Basic Equations and Assumptions}
We solve the two-dimensional, resistive and compressible MHD equations in Cartesian coordinates $(x,z)$, where the z-direction is antiparallel to the gravitational acceleration. 
We assume that the medium is an inviscid perfect gas with a ratio of the specific heats of $\gamma=5/3$. 
We include the radiative cooling effect as described below but neglect the heat conduction (For the discussion on this validity, see section~\ref{cnd_dis}).
The unit of length, velocity, time and density in the simulation are $H_0$, $C_{s0}$, $H_0/C_{s0}\equiv \tau$ and $\rho_0$, respectively, where $H_0=k_{\rm B}T_0/(m g_{0})$ is the pressure scale height ($k_{\rm B}$ is the Boltzmann constant and $m$ is the mean molecular mass), $C_{s0}$ the sound speed and $\rho_0$ the density at the base of the photosphere ($z=0$). 
The gas pressure, temperature, magnetic field strength and energy are normalized by those units, i.e.,  $p_0=\rho_0 C_{s0}^2$, $T_0=mC_{s0}^2/(\gamma k_{\rm B})$, $B_0=(\rho_0 C_{s0}^2)^{1/2}$ and $E_0=\rho_0 C_{s0}^2 H_0^3$, respectively.
The gravity is normalized by $g_0=C_{s0}^2/(\gamma H_0)$.
The values of the normalization units are given as the typical values in the photosphere: $H_0=170$~km, $C_{s0}=6.8$~km~s$^{-1}$, $\tau=H_0/C_{s0}=25$~s and $\rho_0=1.4\times 10^{-7}$~g~cm$^{-3}$. Thus we obtain $p_0=6.3\times 10^4$~dyn~cm$^{-2}$, $T_0=5,600$~K, $B_0=250$~G and $E_0=3.1\times 10^{26}$~erg.

\par
The basic equations are as follows:
\begin{eqnarray}
\frac{\partial \rho}{\partial t}+\rho(\mbox{ \boldmath $v$} \cdot\nabla)&=&-\rho\nabla\cdot\mbox{ \boldmath $v$}\\
\frac{\partial \mbox{ \boldmath $v$}}{\partial t}+(\mbox{ \boldmath $v$}\cdot\nabla)\mbox{ \boldmath $v$}&=&-\frac{1}{\rho}\nabla p+\frac{1}{4\pi \rho}(\nabla\times \mbox{\boldmath $B$})\times \mbox{\boldmath $B$}+\mbox{\boldmath $g$}\\
\frac{\partial T}{\partial t}+(\mbox{\boldmath $v$}\cdot \nabla)T&=&-(\gamma-1)T\nabla\cdot\mbox{\boldmath $v$} -\frac{T}{\tau_{cooling}}\\
p&=&\frac{k_B}{m}\rho T\\
\frac{\partial \mbox{\boldmath $B$}}{\partial t}-c\nabla \times \mbox{\boldmath $E$}&=&\mbox{\boldmath $0$}\\
\mbox{\boldmath $E$}&=&\eta \mbox{\boldmath $J$}-\frac{\mbox{\boldmath $v$}}{c}\times\mbox{\boldmath $B$}\\
\mbox{\boldmath $J$}&=&\frac{c}{4\pi}\nabla \times \mbox{\boldmath $B$}
\end{eqnarray}
Here, $\boldmath{g}=(0,-g_0)$ is the gravitational acceleration, $\eta$ is the resistivity and $\boldmath{J}$ is the current density. 
$\tau_{cooling}$ represents the radiative cooling time given as a function of $z$.
We include the radiative cooling effect in a simple manner: we assume that the radiative cooling time varies linearly in the photosphere and chromosphere and constant in the corona. 
The cooling term is only applied to the regions where the pressure fluctuation is positive compared to the initial pressure (therefore we do not apply the radiative cooling to the expanded regions).
We do not consider the radiative cooling in the convectively unstable zone ($z<0$).
\begin{equation}
\tau_{cooling}(z)= \left \{
\begin{array}{l}
\infty\hspace{10mm}(z<0)\\
(\tau_{\rm c1}-\tau_{\rm c0})z/z_{\rm tr}+\tau_{c0}\hspace{10mm}(0\le z\le z_{\rm tr})\\
\tau_{c1}\hspace{10mm}(z>z_{\rm tr}),
\end{array}
\right.
\end{equation}
where $\tau_{\rm c0}(=0.04\tau_0$=1~s) and $\tau_{\rm c1}(=40\tau_0$=1,000~s) are the radiative cooling time at bottom and top of the chromosphere, respectively. Here we roughly mimic the average cooling time given in VAL-C model \citep{valc}. 
$z_{tr}$ is the height of the transition region.
We also assume the anomalous resistivity model in the simulation box.
\begin{equation}
\eta=\left \{
\begin{array}{l}
0 \hspace{26mm} {\rm for }\hspace{5mm}v_d<v_c \\
\alpha(v_d/v_c-1)^2 \hspace{5mm} {\rm for}\hspace{5mm} v_d\ge v_c
\end{array}
\right.
\end{equation}
\begin{eqnarray}
v_d&=&|\mbox{\boldmath $J$}|/\rho\\
v_c&=&v_{c1}+(v_{c2}-v_{c1})\left\{ \frac{1}{2}\left[ \tanh{ \left( \frac{ z-z_{\rm tr} }{w_c}\right) } +1\right]\right\},
\end{eqnarray}
where $\alpha(=0.01H_0^2/\tau)$, $v_{d}$ and $v_c$ are a constant, the (relative ion-electron) drift velocity and the threshold above which the anomalous resistivity sets in, respectively.
$v_{c1}(=50C_{s0})$ and $v_{c2}(=3,000C_{s0})$ are the threshold in the chromosphere and the corona, respectively and $w_c=0.5H_0$.

\subsection{Initial and Boundary Conditions}
We simply model the solar atmosphere: a super-adiabatically stratified layer representing the upper convection zone ($Z_{\rm min} \le z < 0$), an isothermal cool ($T=T_0$) layer representing the photosphere/chromosphere ($0\le z \le z_{\rm tr}$), and an isothermal hot ($T=150T_0$) layer representing the corona ($z>z_{\rm tr}$).
We take $z=0$ to be the base height of the photosphere.

\par
The initial distribution of temperature is given as
\begin{equation}
T(z)=T_{\rm pho} - \left( a\left| \frac{dT}{dz}\right |_{\rm ad} \right)z \hspace{5mm} (Z_{\rm min}\le z \le 0)
\end{equation}
for convectively unstable zone and
\begin{equation}
T(z)=T_{\rm pho} +(T_{\rm cor}-T_{\rm pho})\left \{ \frac{1}{2}\left[ \tanh{\left( \frac{z-z_{\rm tr}}{w_{\rm tr}} \right)}+1 \right] \right \} \hspace{5mm} (z \ge 0)
\end{equation}
for the upper atmosphere, where $T_{\rm pho}$ and $T_{\rm cor}$ are the temperatures in the photosphere/chromosphere and the corona, respectively.
We took $z_{\rm tr}=10H_0~(1700\rm km)$, $w_{\rm tr}=0.2H_0~(34\rm km)$.
$|dT/dz|_{\rm ad}\equiv (\gamma-1)/\gamma$ is the adiabatic temperature gradient and $a$ is a dimensionless constant.
The layer is convectively unstable in the case that $a>1$. We used $a=1.5$ to model the convection zone.
The two isothermal layers are joined through the transition region with a steep temperature gradient whose width is $\sim 0.4H_0$.
As a result, the transition region physically corresponds to a density contact discontinuous layer.

\par
\begin{figure}
  \begin{center}
    \FigureFile(80mm,80mm){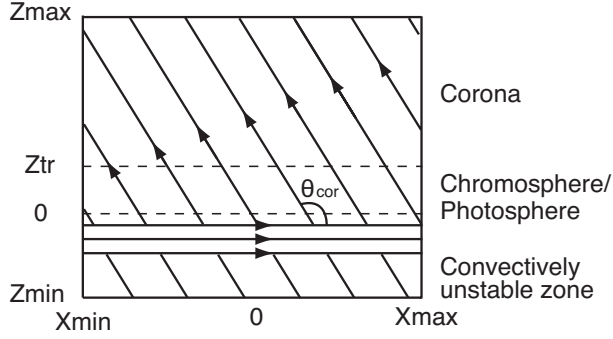}
  \end{center}
  \caption{Schematic picture of the initial condition. Sold lines are the magnetic field lines. The computational box is in the range $-X_{\rm min}\le x \le X_{\rm max}$ and $Z_{\rm min}\le z \le Z_{\rm max}$.}\label{ic}
\end{figure}

For the initial magnetic field we put a magnetic flux sheet in the convectively unstable zone and uniform background field.
The former represents the flux sheet which produces emerging flux and the latter represents the pre-existing magnetic field in the chromosphere and the corona.
A schematic illustration of the model is shown in Figure~\ref{ic}.

The magnetic flux sheet in the convectively unstable zone is given as
\begin{equation}
{\boldmath B}_{\rm fs}(z)=\left[ \frac{8\pi p(z)}{\beta(z)} \right]^{1/2}\hat{\mbox{\boldmath $x$}}
\end{equation}
where
\begin{equation}
\beta(z)=\frac{\beta_{\rm fs}}{f(z)}
\end{equation}
and
\begin{equation}
f(z)=\frac{1}{4}\left[ \tanh{\left(  \frac{z-z_0}{w_0} \right)+1} \right]\left[ -\tanh{\left( \frac{z-z_1}{w_1} \right)} +1\right].
\end{equation}
Here, $\beta_{\rm fs}$ is the plasma beta at the center of the magnetic flux sheet, and $z_0$ and $z_1=z_0+D$ are the heights of the lower and upper boundaries of the flux sheet, respectively, where $D$ is the vertical thickness of the sheet.
We use $z_0=-4H_0$,$D=2H_0$, $\beta_{\rm fs}=5$ and $w_0=w_1=0.2H_0$.
The initial magnetic field strength $B_{\rm fs}$ is $\sim 4B_0(\sim10^3$~G) at the center of the sheet.

\par
Using the initial plasma beta and temperature distribution mentioned above,
we calculate the initial density and pressure distribution numerically using the equation of static pressure balance:
\begin{equation}
\frac{d}{dz}\left[  p(z)+ \frac{B_{\rm fs}^2(z)}{8\pi} \right] +\rho(z){\it g}=0.
\end{equation}

\par
\begin{figure}
  \begin{center}
    \FigureFile(80mm,80mm){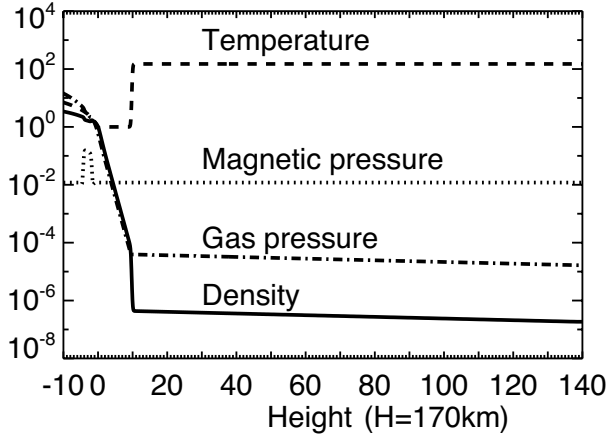}
  \end{center}
  \caption{Vertical distributions of the temperature,magnetic pressure, gas pressure and density in the initial condition. The units are 5,600~K for temperature, $6.3\times10^4$~dyn~cm$^{-2}$ for the gas and magnetic pressure, and $1.4\times10^{-7}$~g~cm$^{-3}$ for the density, respectively.}\label{ic_graph}
\end{figure}

The uniform background field is given as
\begin{eqnarray}
\mbox{ \boldmath $B_{\rm cor}$}=(B_{\rm cor}\cos{\theta_{\rm cor}})\hat{\mbox{\boldmath $x$}}+(B_{\rm cor}\sin{\theta_{\rm cor}})\hat{\mbox{\boldmath $z$}},
\end{eqnarray}
where $B_{\rm cor}$ is the strength of the background field and $\theta_{\rm cor}$ is the angle of the field from the x-axis.
We took $B_{\rm cor}=0.3B_0~(75\rm G)$ and $\theta_{\rm cor}=2\pi/3$.
We show the vertical distribution of physical parameters in the initial condition in Figure~\ref{ic_graph}.

\par
A small-density perturbation of the form
\begin{equation}
\delta \rho=Af(z)\rho(z)\cos{(2\pi x/\lambda)}
\end{equation}
is introduced to the magnetic flux sheet within the finite horizontal domain ($-\lambda/4<x<\lambda/4$), where $\lambda$ is the perturbation wavelength and $A$ is the maximum value of the initial density reduction. 
We took $\lambda=15H_0$ and $A=0.1\rho_0$. 
The wavelength is nearly the most unstable wavelength of the linear Parker instability \citep{parker66}. 

\par
We use periodic boundary conditions in horizontal direction, free boundary conditions at the top, and free boundary conditions for $B_x$ and $B_z$ and symmetric boundary conditions for the other parameters at the bottom.

\par
The size of the simulation box is $-70H_0\le x \le 70H_0$ and $-10H_0\le z \le 140H_0$, namely 23,800~km $\times$ 25,500~km, which is large enough to include the observed chromospheric jets ($\sim 1,000$~km). The grid spacing in horizontal direction is uniform with $\Delta x=0.1H_0$ in $-40H_0\le x \le 40H_0$ and increases upto 1.5$H_0$ in $x<-40H_0$ and $40H_0<x$. In vertical direction $\Delta z=0.05H_0$ in $z\le 30H_0$ and increases up to 1.0$H_0$ in $z>30H_0$. The total grid number is $N_x\times N_y=828\times906$. 

\par
The numerical scheme we adopted is based on the CIP-MOCCT method \citep{kud99}.
The scheme can capture the profiles of physical parameters with contact discontinuities. 
Therefore using this method we can investigate the dynamics of the transition region in detail.
We used some of the routines contained in CANS (Coordinated Astronomical Numerical Software) maintained by Yokoyama et al.

\section{Numerical Results} \label{sec:result}

\subsection{Evolution of Emerging Flux}
\begin{figure}
  \begin{center}
    \FigureFile(80mm,80mm){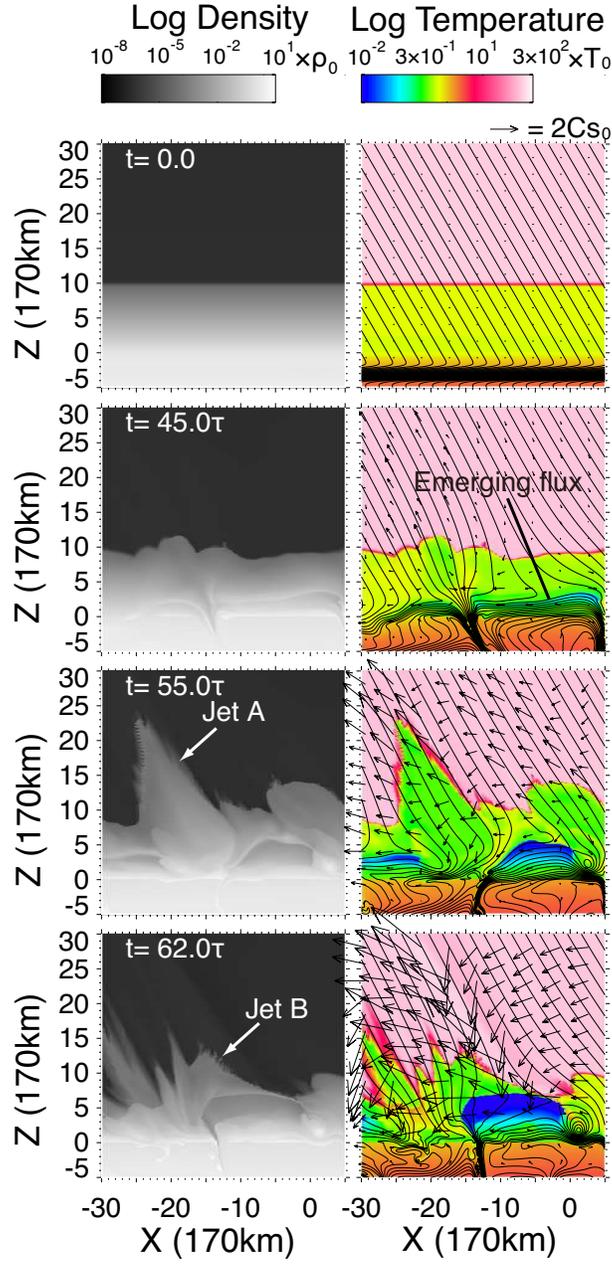}
  \end{center}
  \caption{Time evolution of the density (left) and the temperature (right). The black solid lines show the magnetic field lines. The arrows show the velocity vectors.  $\rho_0=1.4\times10^{-7}$~g~cm$^{-3}$, $T_0=5,600$~K, $C_{s0}=6.8$~km~s$^{-1}$ and $\tau = 25$~s.}\label{evo}
\end{figure}

Magnetic loops expand as a result of the convective Parker instability \citep{parker66, noz92}.
The time evolution of the magnetic loops is shown in Figure~\ref{evo}.
In Figure~\ref{evo}, two jets are indicated by the white arrows (Jet~A and B).
We focus on these jets and investigate how magnetic reconnection creates these jets.

\subsection{Jet Resulting from Magnetic Reconnection near Photosphere: Jet A}\label{sec:bottom}
\begin{figure}
  \begin{center}
    \FigureFile(80mm,80mm){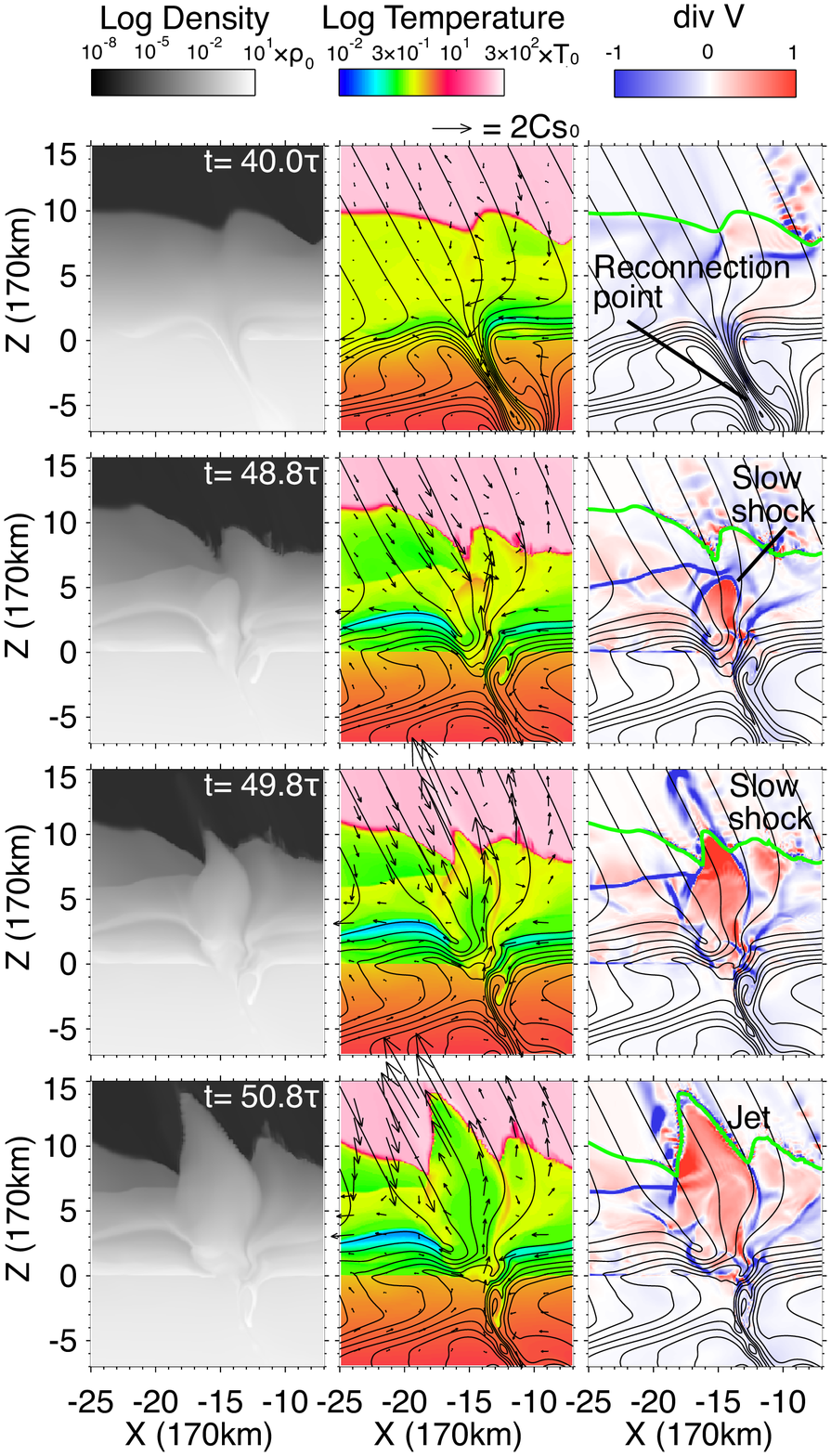}
  \end{center}
  \caption{Jet A: Time evolution of the density (left), the temperature (middle) and the divergence of the velocity field (div V; right). The black solid lines show the magnetic field lines. The arrows show the velocity vectors. The green solid lines show the transition region. Note that in the images of div V the blue and the red regions correspond to the compressed and the expanded regions, respectively. $\rho_0=1.4\times10^{-7}$~g~cm$^{-3}$, $T_0=5,600$~K, $C_{s0}=6.8$~km~s$^{-1}$ and $\tau = 25$~s.}\label{foot}
\end{figure}

The foot-points of the emerging loops sink into the convection zone because the plasma in the lifted flux sheets falls down along the magnetic field and the foot-points become heavy.
The sinking motion naturally creates an antiparallel field (U-shape structure) and a current sheet.
This behavior is also found in the models presented by \citet{iso07} and \citet{arc09}.
As the result of the current sheet formation, magnetic reconnection takes place just below the photosphere ($z\sim -4H_0$) (see Figure~\ref{foot}). 
In our simulation, the numerical resolution is insufficient to resolve the reconnection region or the current sheet.
Although we cannot capture detailed structures of the reconnection region,
we infer that the simulation describes the global dynamics like the acceleration of the reconnection outflow.
The reconnection rate of the numerical reconnection is $\sim$0.1. 
If a fast reconnection takes place near the actual photosphere, the reconnection outflow will generate slow mode waves as found in this simulation.

\par
The reconnection outflow is drastically slowed down by the dense and high-beta photospheric plasma within a short travel distance.
The outflow moves upward along a magnetic field due to both the inertia effect and squeezing.
The upward flow collides with the downward flow along the field.
As a result, a compressed region is formed, leading to the generation of a slow mode wave.
This behavior is also found in the model presented by \citet{tak01}.

\par
The slow mode wave shows wave amplification when it propagates upward through the gravitationally stratified atmosphere as a result of the law of conservation of energy.
Eventually the slow mode wave becomes a slow shock.
This will be analyzed in detail in Section~\ref{sec:discussion}.
The slow shock produces a chromospheric jet through the interaction with the transition region.
Note that to guide the slow mode wave generated by magnetic reconnection in the lower atmosphere, an ambient field plays an important role.
The chromospheric material of the jet is expanded (see div~V distribution of Figure~\ref{foot}) and the density is decreased.
This may be important for disappearance of jets in images of chromospheric lines.

\par
\begin{figure}
  \begin{center}
    \FigureFile(100mm,80mm){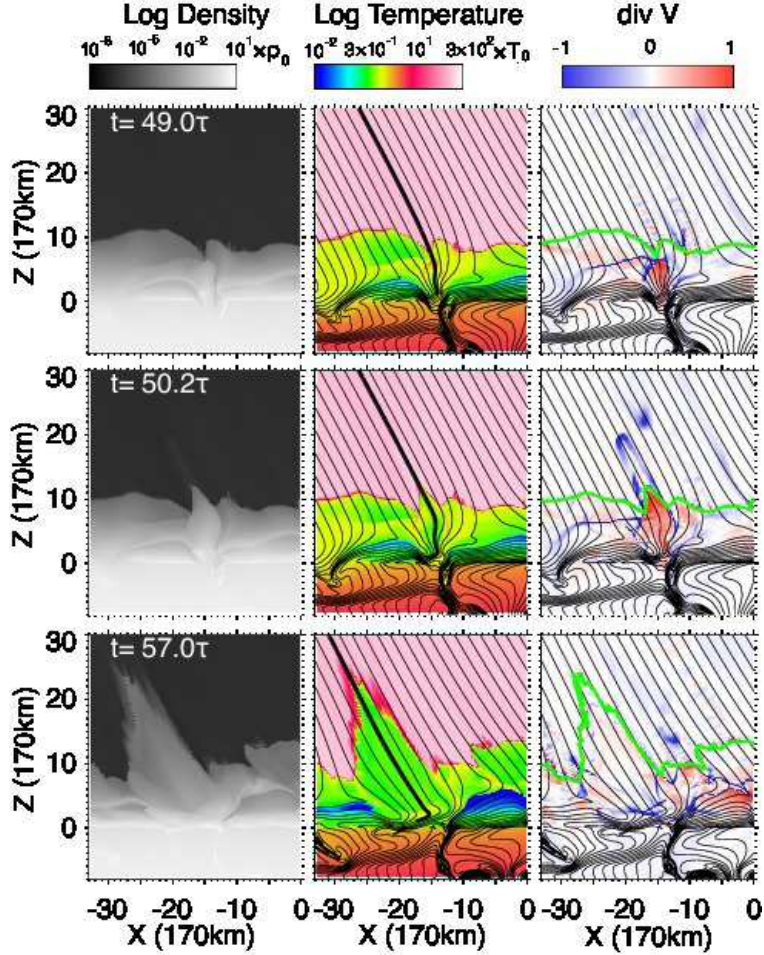}
  \end{center}
  \caption{Jet A: Time evolution of the density (left), the temperature (middle) and the divergence of the velocity field (div V; right). The black solid lines show the magnetic field lines. The data shown in Figure~\ref{td02} were obtained along the thick black lines on the temperature distributions. 
Those thick black lines are the same field line.}\label{line02}
\end{figure}
\begin{figure}
  \begin{center}
    \FigureFile(140mm,60mm){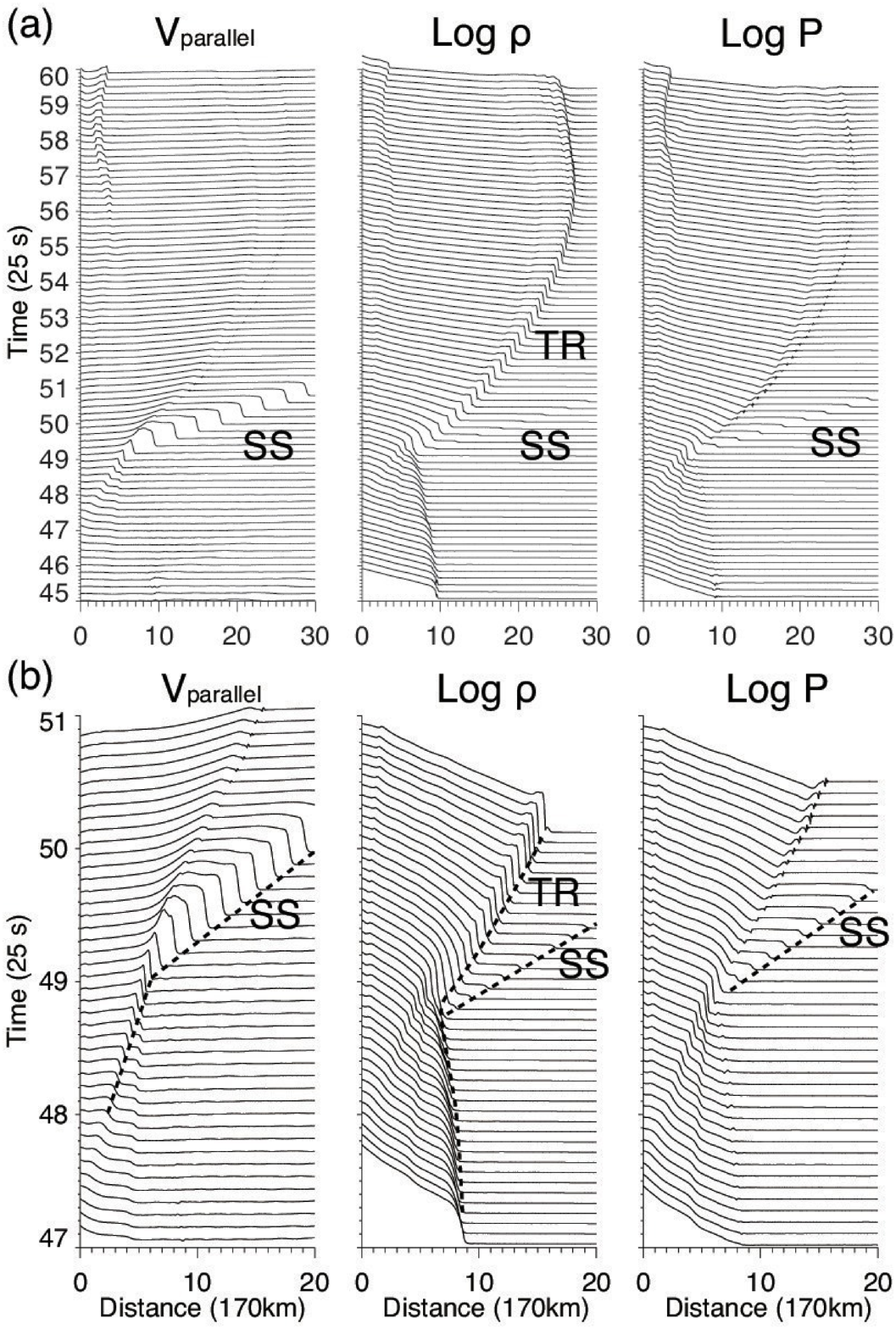}
  \end{center}
  \caption{Jet A: (a) Time evolution of the velocity parallel to the field line, the density and the pressure. These data were obtained along the field line shown in Figure~\ref{line02}. (b) A close-up image of the period between $t=47\tau$ and $51\tau$. SS and TR stand for the slow shock and the transition region, respectively.}\label{td02}
\end{figure}

We obtained the physical parameters along the field line indicated in Figure~\ref{line02}.
Figure~\ref{td02} shows the time evolution of the velocity parallel to the magnetic field line, the density and the pressure measured along the field line.

\par
The growth of the shock, i.e., the growth of the amplitude of the velocity parallel to the field line, can be seen in Figure~\ref{td02} (b).
The velocity of the jet is $\sim6C_{s0}\sim40$~km~s$^{-1}$.
This is much larger than the local sound speed.
 Note that the upward velocity of the reconnection outflow in the photosphere is $0.6C_{s0}\sim4$~km~s$^{-1}$.
The maximum height of the jet is $\sim 25H_0\sim4,300$~km from the photosphere.
The difference between the maximum height and the initial height is $\sim 17H_0\sim 2,900$~km~s$^{-1}$.

\par
This example shows that even if magnetic reconnection takes place below the photosphere, tall ($\sim 4$~Mm) jets can be produced through the interaction between a slow shock and the transition region.
It should be noted that the jet is not the reconnection outflow but the transition region lifted by the slow shock.

\subsection{Jet Resulting from Magnetic Reconnection in Upper Chromosphere: Jet B}\label{sec:upper}
Magnetic loops push an ambient field and start to reconnect with it as magnetic loops expand.
Chromospheric material is accelerated after magnetic reconnection.
The accelerated chromospheric plasma eventually start to rise upward along the magnetic field, becoming a chromospheric jet.
No hot jet driven by the pressure gradient as in \citet{yok96} coexists with the chromospheric jet.

\par
\begin{figure}
  \begin{center}
    \FigureFile(140mm,80mm){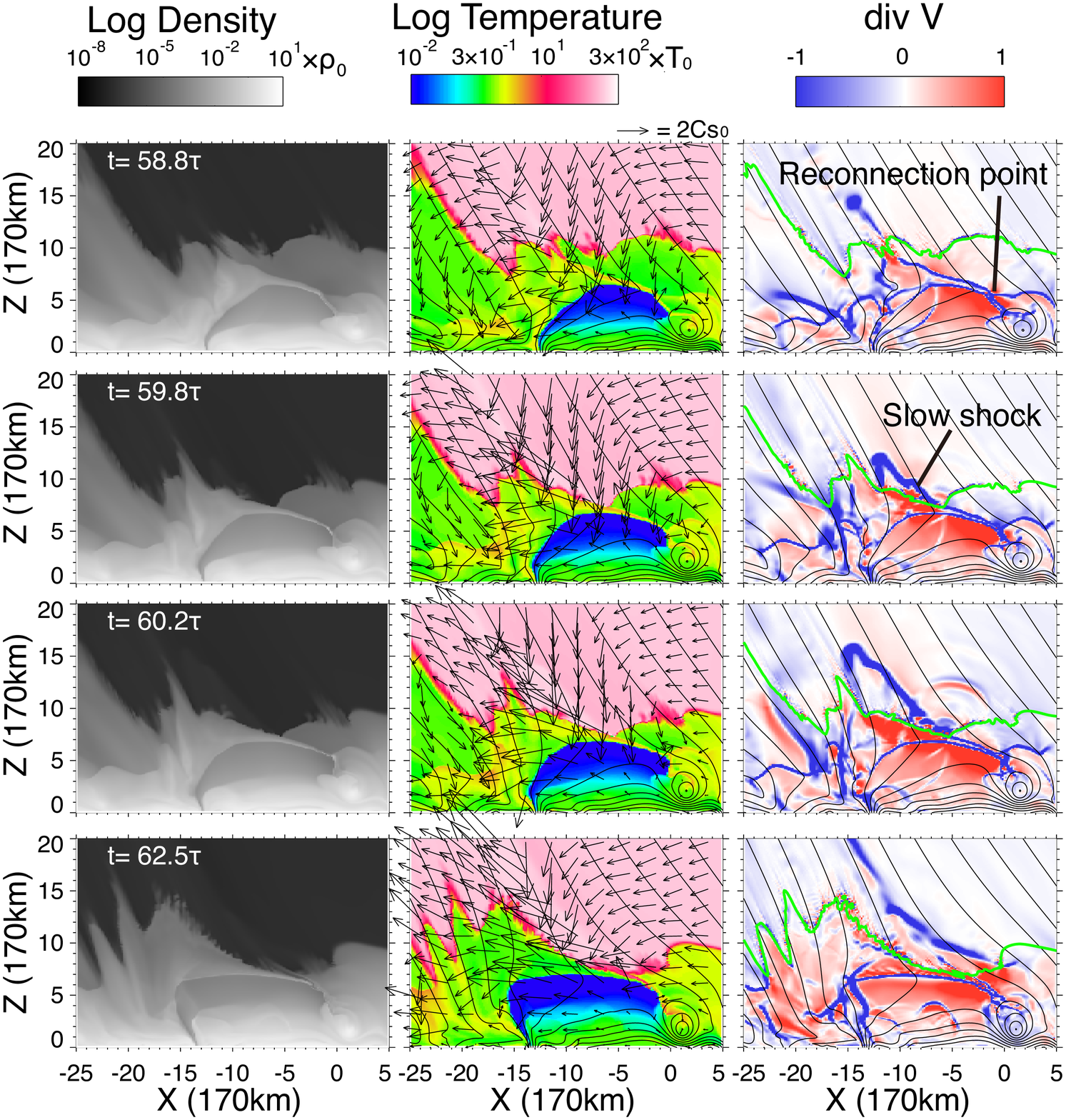}
  \end{center}
  \caption{Jet B: Time evolution of the density (left), the temperature (middle) and the divergence of the velocity field (div V; right). 
  	The black solid lines show the magnetic field lines. The arrows show the velocity vectors. 
	The green solid lines show the transition region. 
	Note that in the images of div V the blue and the red regions correspond to the compressed and the expanded regions, respectively. 
	$\rho_0=1.4\times10^{-7}$~g~cm$^{-3}$, $T_0=5,600$~K, $C_{s0}=6.8$~km~s$^{-1}$ and $\tau = 25$~s.}\label{petschek}
\end{figure}

Figure~\ref{petschek} shows the time evolution of the density, temperature and divergence of the velocity field of the emerging flux region.
The density distributions show a shape similar to that of observed jets, i.e. a jet with a loop structure (see Figure~\ref{overview}).

\par
We assumed the anomalous resistivity model in the simulation (the justification of the anomalous resistivity model will be discussed in Section~\ref{sec:discussion}). 
As a result, the Petschek-type reconnection, in which the standing slow shocks emanate from the magnetic reconnection point, occurs between the emerging flux and the pre-existing ambient field.
In the following, we call the standing slow shocks emanating from the reconnection point the Petschek slow shocks.

\par
\begin{figure}
  \begin{center}
    \FigureFile(80mm,80mm){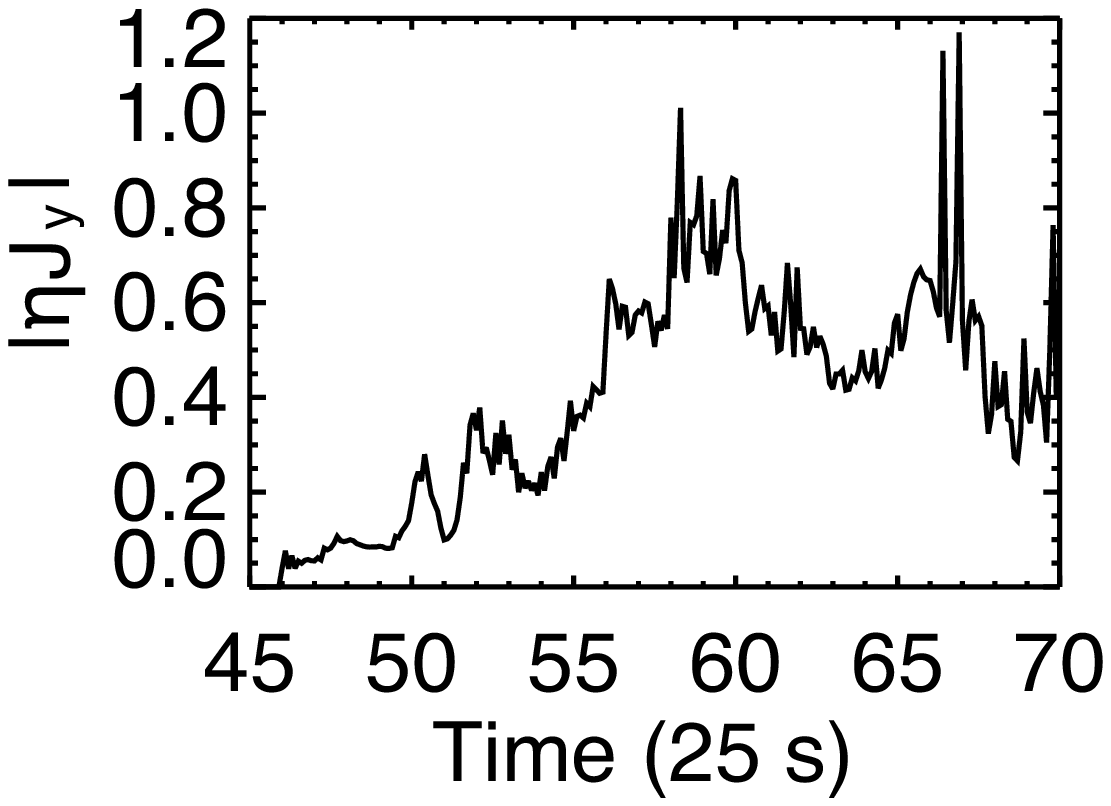}
  \end{center}
  \caption{Jet B: Time evolution of the resistive electric current $|\eta J_{y}|$ at the neutral point as a measure of the reconnection rate.
This is measured in the region $-10H_0< x < 5H_0$ and $H_0< z <10H_0$.
The unit of the time is 25~s. }\label{rxrate}
\end{figure}

Figure~\ref{rxrate} shows the time evolution of the resistive electric current $|\eta J_{y}|$ at the neutral point as a measure of the reconnection rate, where the reconnection rate is defined as the reconnected magnetic flux per unit time.
This is measured in the region $-10H_0< x < 5H_0$ and $H_0< z <10H_0$.
As shown, magnetic reconnection starts at $t\sim 46\tau$.

\par
\begin{figure}
  \begin{center}
    \FigureFile(140mm,80mm){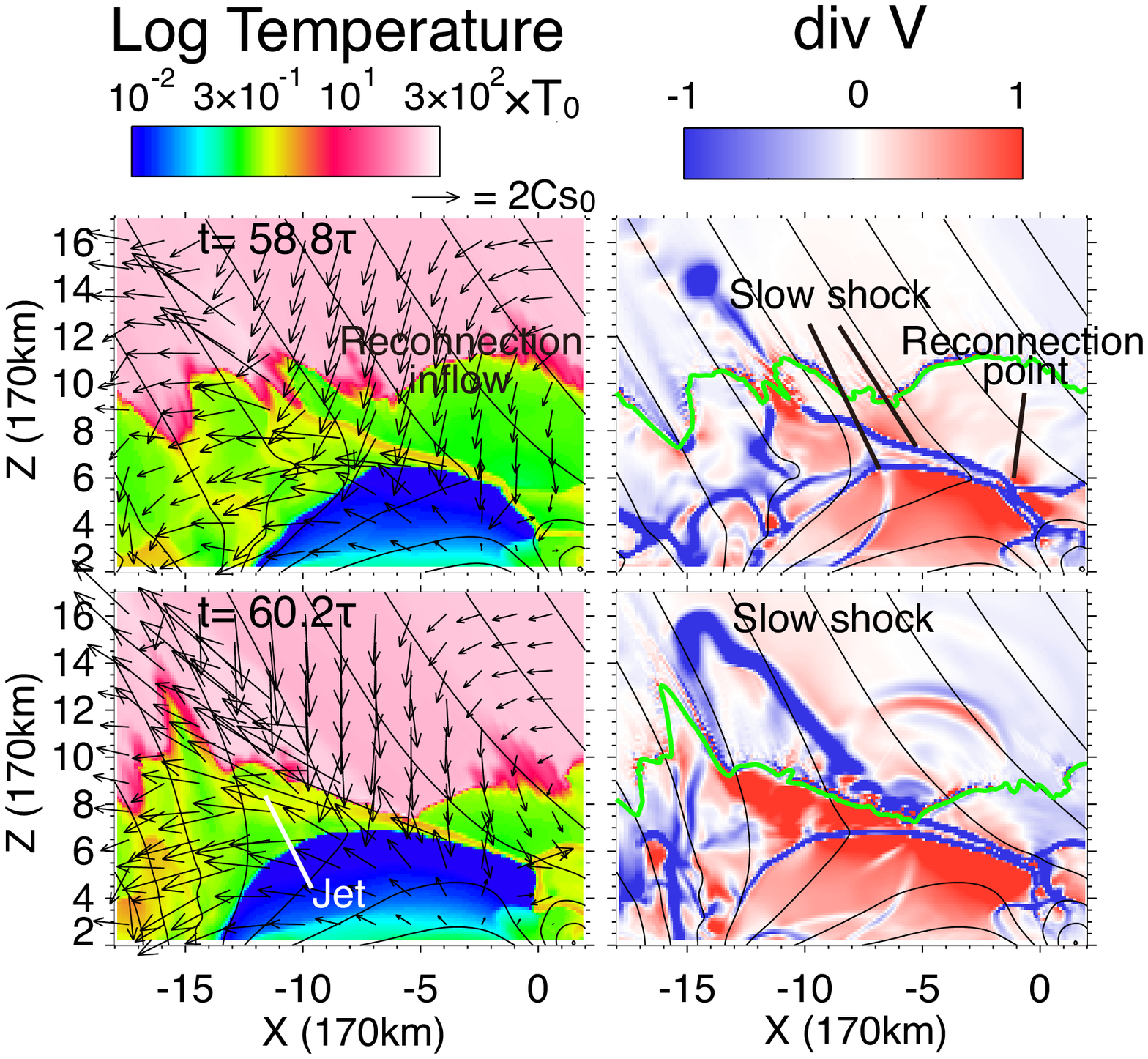}
  \end{center}
  \caption{Jet B: Enlarged images of the temperature (left) and the divergence of the velocity field (right) distributions before (top) and after (bottom) the slow shock crosses the transition region. The green line shows the transition region. $T_0=5,600$~K, $C_{s0}=6.8$~km~s$^{-1}$ and $\tau = 25$~s. }\label{pickup}
\end{figure}

Many shocks are formed in the chromosphere.
The blue and red regions in div~V distribution in Figure~\ref{petschek} correspond to the compressed and expanded regions, respectively.
The shock structures can be discerned as sharp blue regions in Figure~\ref{petschek} (blue regions emanating from the reconnection point are the Petschek slow shocks).
The enlarged images before and after one of the Petschek slow shocks crosses the transition region are shown in Figure~\ref{pickup}.

\par
\begin{figure}
  \begin{center}
    \FigureFile(140mm,80mm){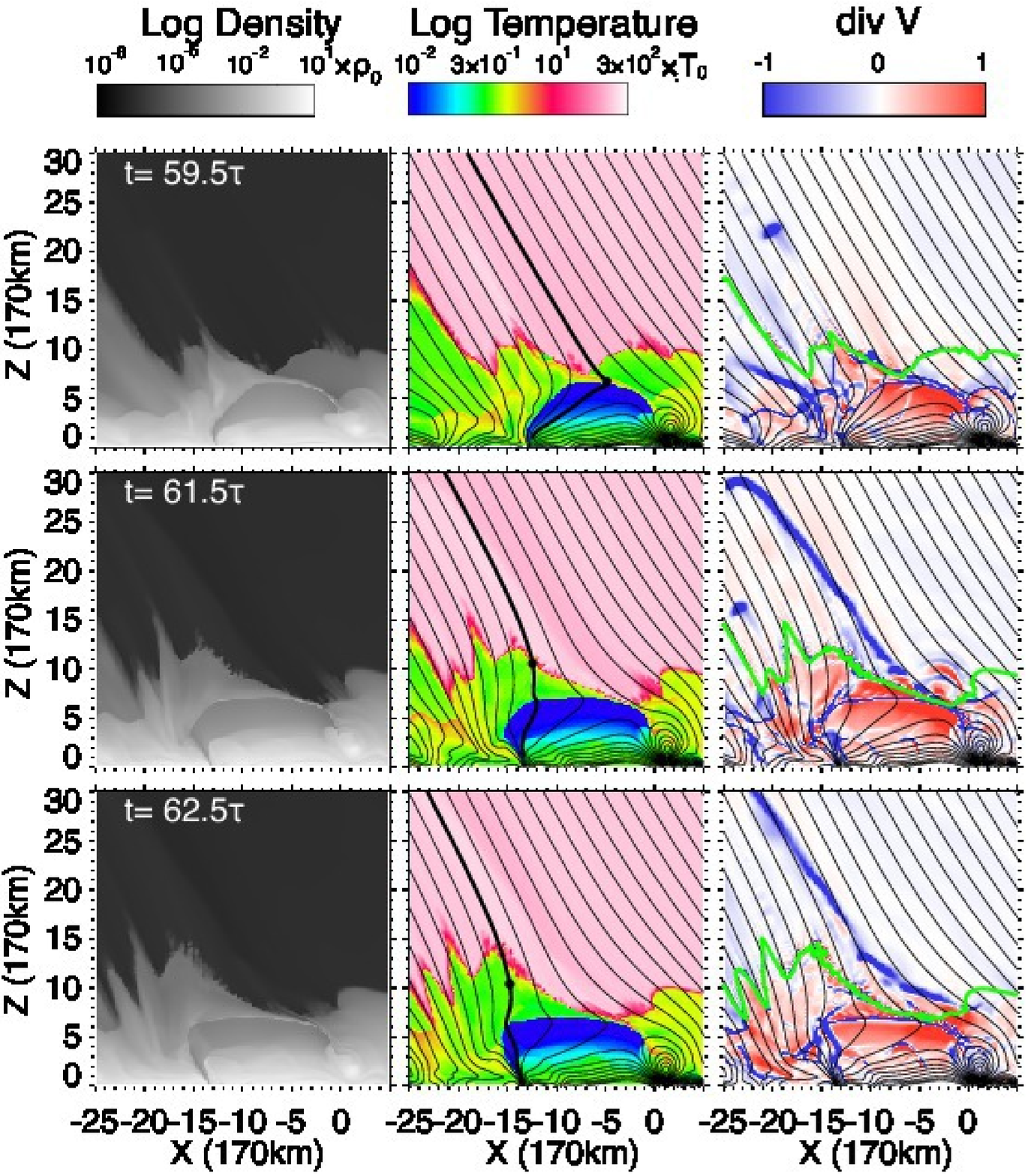}
  \end{center}
  \caption{Jet B: Time evolution of the density (left), the temperature (middle) and the divergence of the velocity field (div V; right). 
  	The black solid lines show the magnetic field lines. The data shown in Figure~\ref{td01} were obtained 
	along the thick black lines on the temperature distributions. 
	Those thick black lines are the same field line.
	The position where the distance of $10H_0$ from the start point is marked with the black point.}\label{line01}
\end{figure}
\begin{figure}
  \begin{center}
    \FigureFile(160mm,160mm){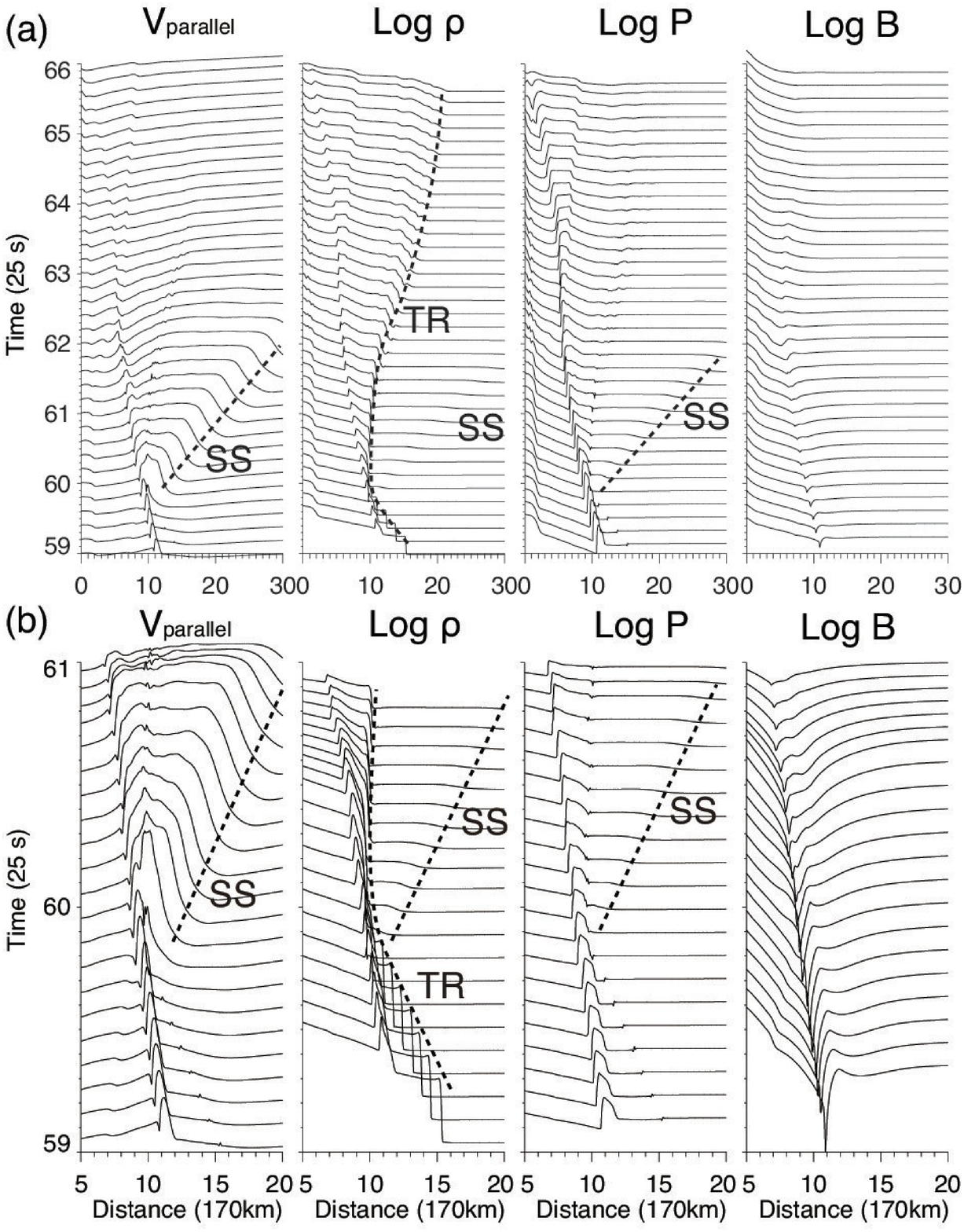}
  \end{center}
  \caption{Jet A: (a) Time evolution of the velocity parallel to the field line, the density and the pressure. These data were obtained along the field line shown in Figure~\ref{line01}. (b) A close-up image of the period between $t=59\tau$ and $61\tau$. SS and TR stand for the slow shock and the transition region, respectively.}\label{td01}
\end{figure}

We obtained the physical parameters along the field line shown in Figure~\ref{line01}.
The black thick field lines are the same field line.
Figure~\ref{td01} shows the evolution of the velocity parallel to the magnetic field line, the density and the pressure measured along the field line.
The enhanced density and the pressure in the outflow region appear as the peaks at distance of about 10$H_0$ in Figure~\ref{td01} (The positions where the distance is $10H_0$ are marked with the black points on the field line in Figure~\ref{line01}).
We can see that the height of the transition region (i.e. the steep gradient in the density profile at $15H_0$ distance from the start point of the line at $t=59\tau$) decreases with time during the period between $t=59\tau$ and 60$\tau$.
This is because the reconnection inflow reduces the height of the transition region (see Figure~\ref{petschek} and \ref{pickup}).
Eventually at $t\sim60\tau$ the transition region crosses one of the Petschek slow shocks attached to the reconnection point.

\par
After reconnection started, the inflow and outflow regions 
are separated by stationary Petschek slow shocks.
We express such a situation in words
gthe slow shock is attached to the outflow region.h
When one of the Petschek slow shocks crosses the transition region, it starts to propagate upward.
Then the shock is no longer attached to the chromospheric material accelerated 
due to magnetic tension (see the bottom panels of Figure~\ref{pickup})
(The propagating speed of a slow shock in the low-beta region is almost the same as the local sound speed, which is much larger in the corona than that in the chromosphere.)
In div~V distributions at $t=59.8\tau$ and later times in Figure~\ref{petschek}, we can see the slow shock (blue region) propagating upward.
After the interaction, we can find that the steep gradient in the magnetic field strength distribution is smoothed out.

\par
\begin{figure}
  \begin{center}
    \FigureFile(90mm,80mm){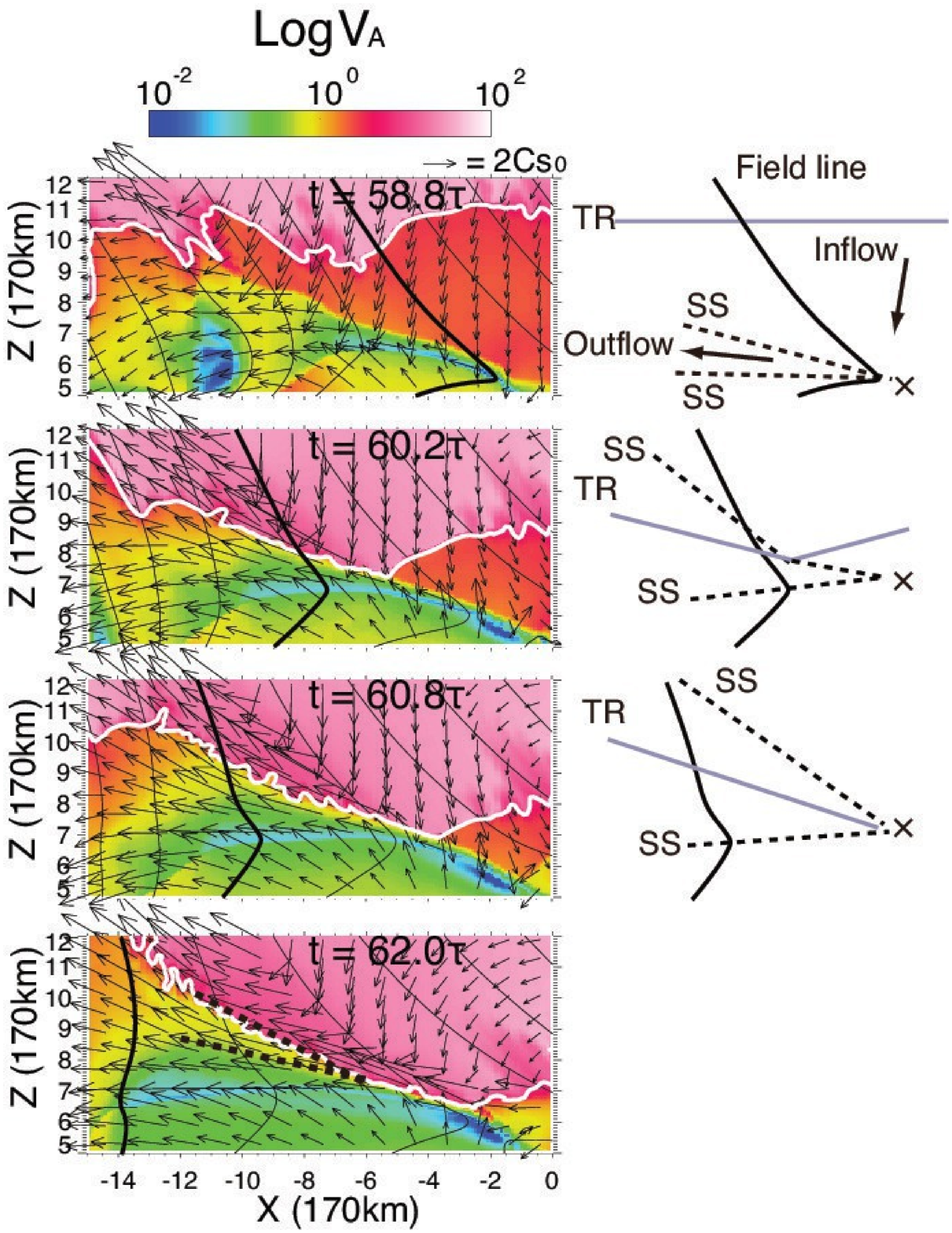}
  \end{center}
  \caption{Jet B: Left: Time evolution of the distribution of the Alfv\'en speed before ($t=58.8\tau$) and after ($t=60.2\tau$, $t=60.8\tau$ and $62.0\tau $) the interaction between the slow shock and the transition region. 
The thick white line shows the transition region. The thick black line show the same filed line. The Alfv\'en velocity in the region between the dashed lines is increased. $C_{s0}=6.8$~km~s$^{-1}$ and $\tau = 25$~s. Right: Schematic diagrams to describe how the slow shock crosses the transition region. SS and TR stand for the slow shock and the transition region, respectively. The cross mark represents the reconnection point.}\label{va2d}
\end{figure}

After the interaction between the Petschek slow shock and the transition region, the reconnection outflow, which is almost in the horizontal direction, is accelerated, although the transition region is not lifted upward yet (see Figure~\ref{td01} (b)).
Considering that the outflow speed is comparable to the local Alfv\'en speed, it is worth showing the Alfv\'en speed distribution of the reconnection region.
Figure~\ref{va2d} shows that the Alfv\'en speed in the outflow region becomes large after the interaction.
In the region where the Alfv\'en speed increases, the outflow velocity also has increased.

\par
To investigate the interaction process in detail, we performed 1.5D MHD simulations and analyzed the interaction between one of Petschek slow shocks and the transition region.
We summarize the simulation results in Appendix.
We found that the outflow is further accelerated due to magnetic tension after the interaction between the Petschek slow shock and the transition region.
When the slow shock crosses the transition region, a slow mode rarefaction wave is generated and then propagates into the outflow region.
The slow mode rarefaction wave causes the density and the magnetic field strength to become low and high, respectively.
Thus the Alfv\'en speed becomes large (see Figure~\ref{va2d}).

\par
When magnetic reconnection takes place, a magnetic field tends to be relaxed and re-configured to a relatively straight structure. Reconnection outflows are accelerated as a consequence of the relaxation process. 
The Alfv\'en velocity near the outflow region determines the outflow speeds. 
Therefore,
the relaxation process will proceed quickly when the Alfv\'en velocity increases.
Therefore the high Alfv\'en speed leads to quick relaxation of the configuration of the magnetic field.
The re-configuration process results in the further acceleration of the reconnection outflow.

\par
\begin{figure}
  \begin{center}
    \FigureFile(80mm,80mm){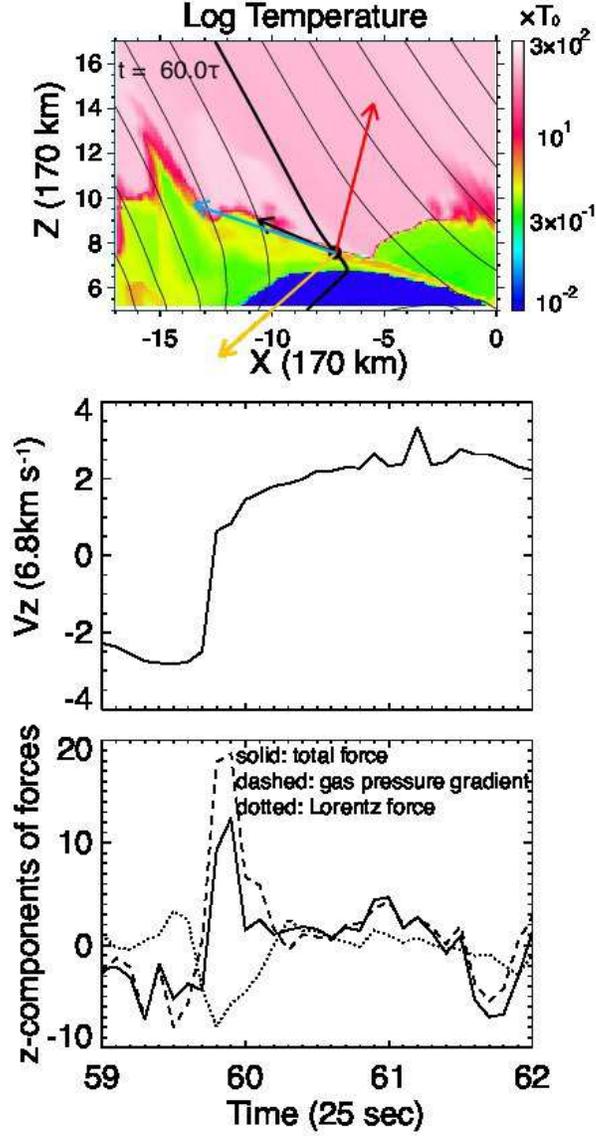}
  \end{center}
  \caption{Jet B: (top) Temperature distribution with the velocity and three force vectors. The velocity and the forces are measured just below the transition region along the field line in Figure~\ref{line01}. (middle and bottom) Time evolution of the z-components of the velocity and the forces. The unit of the forces per unit mass $F_0$ is $C_{s0}/\tau=0.27$~km~s$^{-2}$, where $C_{s0}=6.8$~km~s$^{-1}$ and $\tau = 25$~s.}\label{force}
\end{figure}

How is the chromospheric plasma accelerated upward due to magnetic reconnection?
We noted that the upward motion of the transition region starts around at $t=62\tau$.
To clarify how the jet is accelerated, we show the forces acting on the chromospheric plasma just below the transition region in Figure~\ref{force}.
The plasma in the outflow region feels no strong upward acceleration during $t=60\tau$ and $62\tau$.
The z-component of the velocity shows the gradual acceleration during the period.
Note that the chromospheric plasma in the outflow feels no strong acceleration due to the pressure gradient force when the outflow collides with the ambient field.
The chromospheric plasma (reconnection outflow) which is almost horizontally accelerated eventually starts to rise upward along the magnetic field after colliding with the ambient field.

\par
\begin{figure}
  \begin{center}
    \FigureFile(80mm,80mm){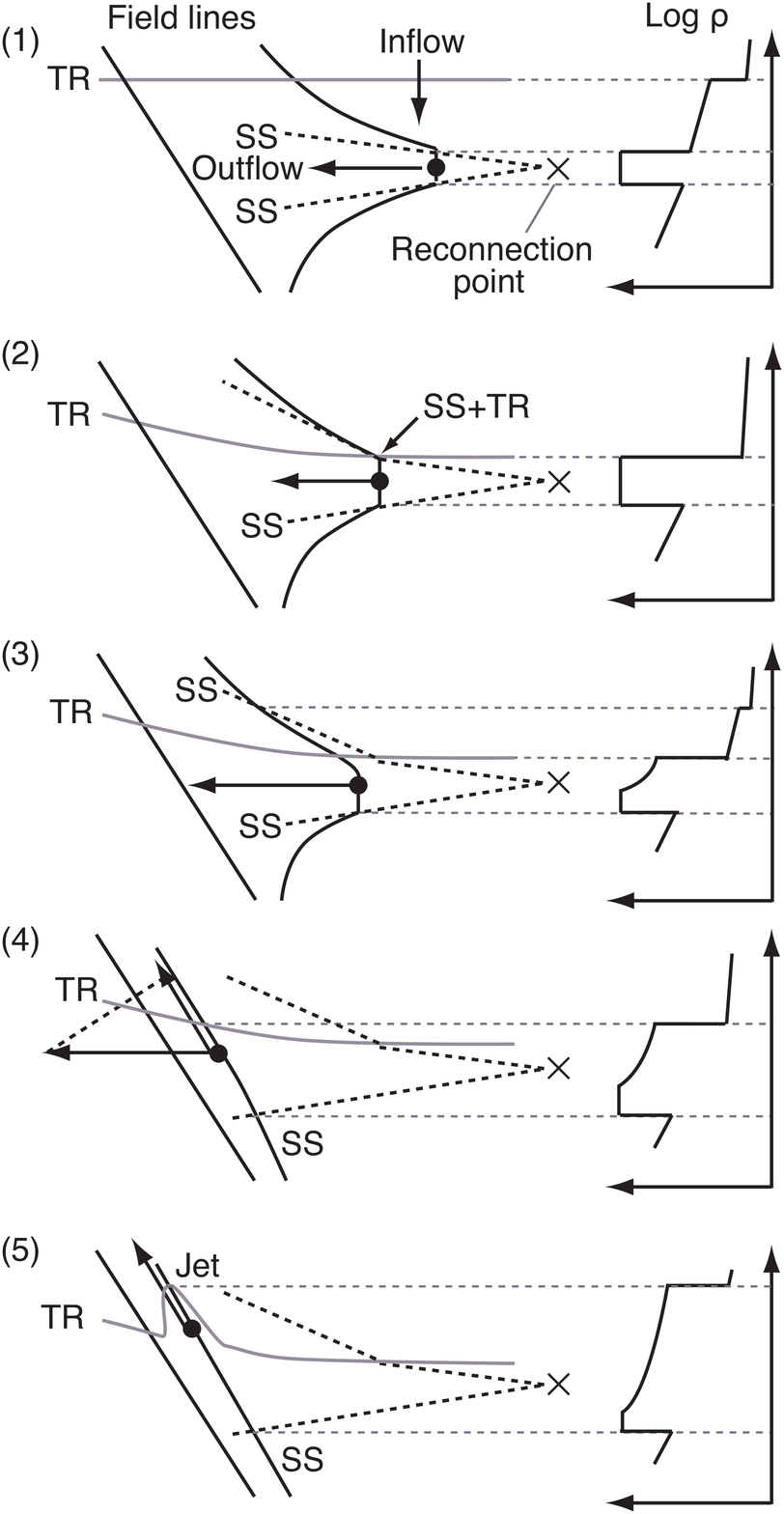}
  \end{center}
  \caption{Jet B: Schematic diagram to describes how the the transition region starts to rise upward ("whip-like acceleration"). TR and SS stand for the transition region and slow shock, respectively. For a detailed discussion on the SS-TR interaction, see Appendix. }\label{col_ponchi}
\end{figure}

A schematic diagram of the acceleration process is shown in Figure~\ref{col_ponchi}.
After the outflow collides with the ambient field, the outflow plasma moves in the direction along the ambient field.
Consequently, we obtain the chromospheric jet.
This process which we call the "whip-like acceleration" is essentially the centrifugal force acceleration.

\par
The upward velocity ($V_z$) of the chromospheric jet is $\sim 2.5C_{s0}\sim 17$~km~s$^{-1}$.
The maximum height of the jet is $\sim15H_0\sim2,600$~km from the base of the photosphere (Figure~\ref{td01}).
The difference between the maximum height and the initial height is $\sim 8H_0\sim 1,400$~km~s$^{-1}$.

\par
We can discern the transition region corrugated at the top of the jet (see Figure~\ref{petschek} and \ref{va2d}).
\citet{sto95} found that parallel slow shocks are unconditionally unstable to the corrugation instability.
In out simulation, the slow shock crosses the transition region and propagates almost parallel to the magnetic field.
Therefore the corrugated structure can be a result of the corrugation instability.

\section{Discussion}\label{sec:discussion}
We studied the acceleration mechanisms of chromospheric jets associated with emerging flux using a 2D MHD simulation.
It is found that the height of magnetic reconnection determines the acceleration mechanism of chromospheric jets.
Here we present a schematic diagram summarizing the acceleration mechanisms of chromospheric jets.
We also give implications for observations and discuss the validity of the adopted assumptions.

\subsection{Comparison between One-Dimensional Case and Two-Dimensional Case: Jet A}
We found that a slow mode wave grows to be a shock in the case that magnetic reconnection occurred just below the photosphere.
The growth of the shock in one-dimensional cases has been studied by many authors \citep{ost61,shi82,heg07}.
We investigate how our 2D simulation results differ from the prediction of the idealized 1D theory in which a flux tube is rigid enough not to be disturbed by the plasma dynamics in it and the total energy is exactly conserved in a flux tube (so-called zero-beta approximation).

\par
Here we review the 1D theory. 
We consider the situation in which a slow mode wave propagates upward in a rigid flux tube in a gravitationally stratified atmosphere.
In the linear regime, we obtain the following relation from the energy conservation of the slow mode wave,
\begin{equation}
A\rho v_{\parallel}^2C_s={\rm constant},
\end{equation}
where $A$, $v_{\parallel}$ and $C_s$ are the cross-section of the flux tube, the velocity parallel to the magnetic field and the sound speed, respectively.
Considering the magnetic flux conservation $BA$=constant and that $C_s$ is almost constant in the chromosphere, we obtain
\begin{equation}
v_{\parallel}B^{-0.5}\propto \rho^{-0.5} \label{eq:linear},
\end{equation}
where $B$ is the magnetic field strength.
In the non-linear regime, this relation is modified due to the dissipation processes.
\citet{ono60} derived the relation in a strong shock case:
\begin{equation}
v_{\parallel}B^{-0.5}\propto \rho^{-0.236} \label{eq:nonlinear}
\end{equation}
Note that these relations will hold if the energy is exactly conserved in a flux tube.

\par
\begin{figure}
  \begin{center}
    \FigureFile(70mm,80mm){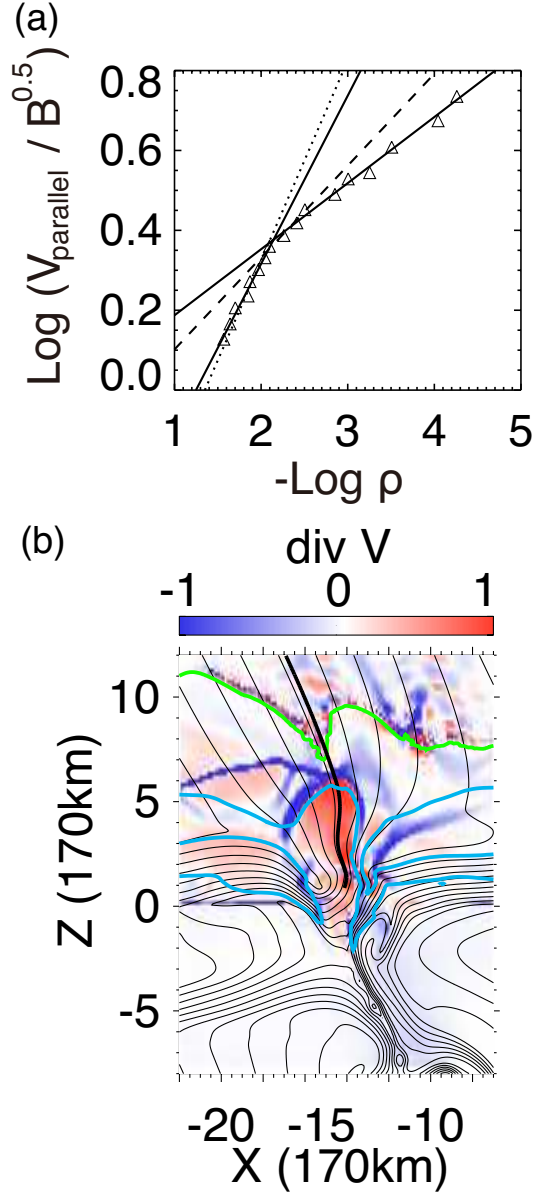}
  \end{center}
  \caption{Jet A: (a) The growth of the slow shock discussed in Section~\ref{sec:bottom}. Triangles show the numerical results. The two solid lines shows the fitting lines calculated using the least square method (one is obtained using left 8 points of the numerical data and the other is obtained using the rest points). The dotted and dashed lines correspond to the analytic relations (\ref{eq:linear}) and (\ref{eq:nonlinear}), respectively. (b) The distribution of the divergence of the velocity field (div V). The black lines are magnetic field lines. The thick black line is the line along which the data in (a) were obtained. The light blue lines correspond to the region where the plasma beta is unity. The green line corresponds to the transition region.}\label{growth}
\end{figure}

We investigate the dependence of the growth rate $v_{\parallel}B^{-0.5}$ on the density $\rho$ using our simulation results.
We show the result in Figure~\ref{growth} (a) (This figure corresponds to Figure~6 in \citet{ss82}.)
In Figure~\ref{growth} (a), the triangles are obtained from the numerical simulation. 
The dependence of the growth rate on the density changes approximately at $-\log{\rho}=2$.
We separately applied the least square method to the numerical results below and above this density value to obtain the fitting lines. These lines correspond to the two solid lines.
These lines have slopes of 0.36 and 0.15, respectively.
The dotted and the dashed lines correspond to the analytic relations~(\ref{eq:linear}) and (\ref{eq:nonlinear}), respectively.

\par
Below $-\log{\rho}\sim2$, the dependence of the growth rate on the density slightly deviates from the prediction of the idealized 1D theory (\ref{eq:linear}).
This is probably because the plasma beta is greater than unity in the low chromosphere and the low-beta approximation is broken.
This means that the energy of the shock wave in a flux tube leaks to other flux tubes.
Above $-\log{\rho}\sim2$, the discrepancy between the numerically obtained relation and the 1D analytic relation~(\ref{eq:nonlinear}) becomes large.
The energy loss of the wave along a flux tube is larger than that predicted with the aid of the one-dimensional analysis.
This implies that the one-dimensional approximation, i.e., the low-beta approximation is broken in the non-linear regime.

\par
To clarify the cause of the discrepancy, we investigate whether the assumption of the idealized 1D theory are valid.
We plot the lines on which the plasma beta is unity in Figure~\ref{growth}~(b).
It should be noted that the plasma beta behind the shock is close to unity even in the upper chromosphere.
Therefore the 1D approximation (zero-beta approximation) is severely broken in the regime and the energy of the slow shock is carried to other flux tubes by fast mode waves.
If we consider the energy conversion rate from the magnetic energy liberated by magnetic reconnection into the kinetic energy of jets, we need pay an attention to this effect.

\subsection{Acceleration of Reconnection Outflow: Jet B}
The reconnection outflow discussed in Section~\ref{sec:upper} is basically accelerated due to the whip-like motion of the reconnected magnetic field.
The reconnection outflow is accelerated after the interaction between one of the Petschek slow shocks and the transition region.
Here we focus on this point, which has not been discussed in the previous studies.

\par
We obtain the slow mode rarefaction wave propagating into the outflow region after the Petschek slow shock crosses the transition region.
Behind the slow mode rarefaction wave, the magnetic field strength increases and the density decreases (see Figure~\ref{td01}).
Thus the Alfv\'en speed increases, as shown in Figure~\ref{va2d}.
The large Alfv\'en speed leads to the rapid re-configuration of the magnetic field.
The re-configuration results in the further acceleration of the outflow plasma.
For a detailed discussion, see Appendix.

\subsection{Schematic Diagram of Acceleration Mechanisms of Jets}\label{sec:diagram}
In the case of the jet associated with magnetic reconnection near the photosphere (Jet~A), the slow mode wave generated as a result of the collision of the reconnection outflow is amplified when it propagates upward. 
The slow mode wave becomes a shock as a result of the law of conservation of energy.
Through the interaction between the slow shock and the transition region, a chromospheric jet is produced.
This is essentially the same process as that discussed by \citet{shi82}, that is, a pure hydrodynamic process.

\par
In the case of the jet associated with magnetic reconnection in the upper chromosphere (Jet~B), the chromospheric plasma (reconnection outflow) is accelerated due ti the whip-like motion of the reconnected magnetic field.
The outflow is further accelerated through the interaction between the Petschek slow shock and the transition region.
This is an MHD effect which cannot be found in the pure hydrodynamic cases (See Section~\ref{sec:upper} and Appendix).

\par
\begin{figure}
  \begin{center}
    \FigureFile(160mm,80mm){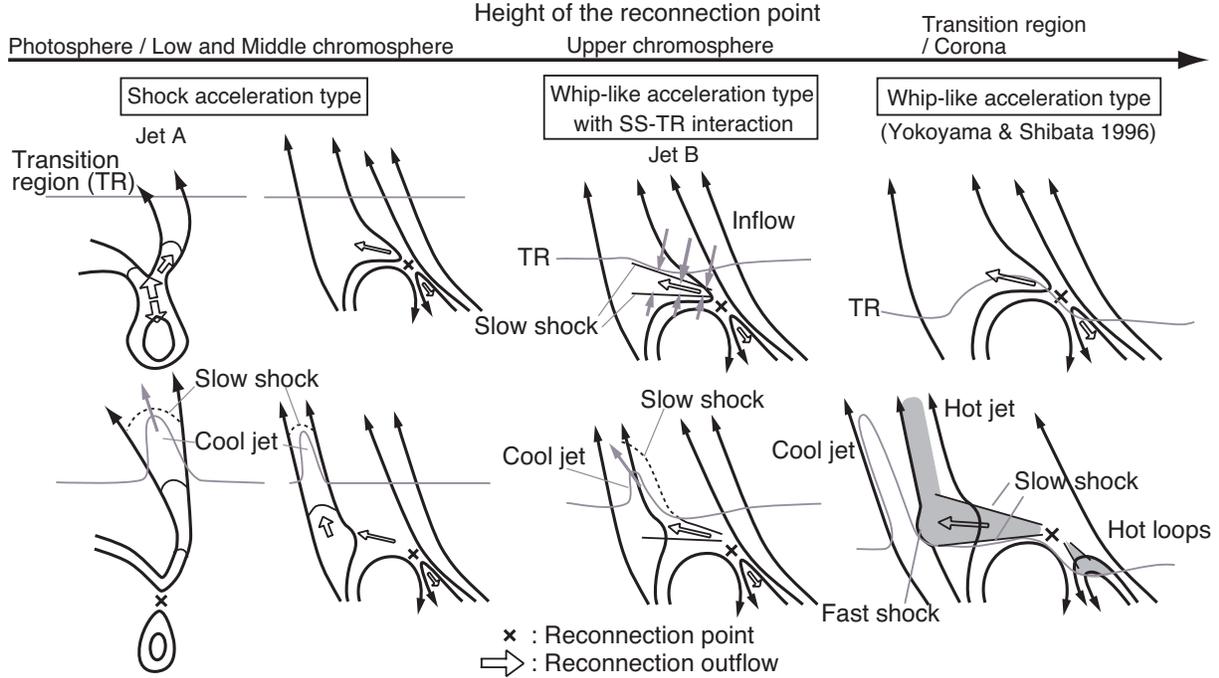}
  \end{center}
  \caption{Schematic diagrams of the generation mechanisms of the chromospheric (cool) jet. The speculated scenario for the low and middle chromospheric cases (the second scenario from the left) and the model found by \citet{yok96} are also included. Jet A and B are the jets discussed in Section~\ref{sec:bottom} and \ref{sec:upper}, respectively.}\label{summary}
\end{figure}

We self-consistently related the acceleration mechanisms of chromospheric jets with magnetic reconnection under the assumptions adopted in this study.
The height of a reconnection point is found to determine the acceleration scenario.
The acceleration processes are summarized in Figure~\ref{summary}.
We include the model found by \citet{yok96} to unify the acceleration processes and the speculated scenario for the low and middle chromospheric cases.

\subsection{Implication for Observations}

\subsubsection{Implication for recurrent Acceleration of Jets}
Observation found the recurrent acceleration of chromospheric anemone jets \citep{nis11}.
Here we discuss a possible recurrent acceleration scenario in terms of the process found in this study.

\par
We showed that slow mode waves can be generated when reconnection outflows collide with an ambient field.
Let us assume that the velocity of the reconnection outflows which hit the ambient field is time-dependent.
This is possible, e.g., if the plasmoids are formed in the current sheet, where the plasmoids are the regions of magnetically confined plasma.
The emergence of the plsasmoids has been theoretically and observationally implied \citep{sin11,lea12,sin12}.

\par
\begin{figure}
  \begin{center}
    \FigureFile(80mm,80mm){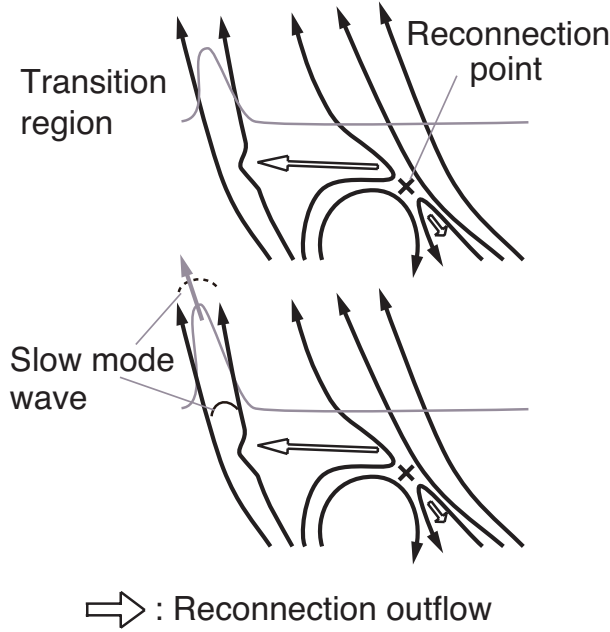}
  \end{center}
  \caption{Schematic diagram of a recurrent acceleration process of the chromospheric jet as a result of the collision of the reconnection outflow. The slow mode waves are generated as a result of squeezing of the plasma at the termination region of the reconnection outflow (a mode conversion.)}\label{reccurent}
\end{figure}

When the outflow hits the ambient magnetic field, slows mode waves will be generated by squeezing.
This process is essentially the same as a mode conversion process from fast mode waves into slow mode waves \citep[e.g.][]{bog03}.
The slow mode waves (or slow shocks) will propagate upward and accelerate jets when passing through the transition region.
\citet{yok96} discussed the direct acceleration due to the pressure enhancement due to the squeezing of the chromospheric plasma by reconnection outflows.
Here we note that slow mode waves can be generated as a result of the collision of reconnection outflows as in \citet{tak01}.
This process will be preferable for the recurrent acceleration of chromospheric jets.
The schematic diagram is shown in Figure~\ref{reccurent}.

\subsubsection{Relation between Ellerman bombs and Surges}
Observations have suggested that the Ellerman bombs result from the emergence of the undulatory flux tubes \citep{pariat04}. 
\citet{wat11} observationally found the morphological evidence for the magnetic reconnection in the deep photosphere.
They also found that some Ellerman bombs were accompanied by dark H$\alpha$ surges.
Following their study, the apparent velocity of the ejections in the photosphere is $\sim 10$~km~s$^{-1}$ and the Doppler velocity of the surge is $\sim40$~km~s$^{-1}$.

\par
We obtained the tall ($\sim 4$~Mm) jet resulting from the reconnection just below the photosphere.
The reconnection outflow velocity in the photosphere is $0.6C_{s0}\sim 4$~km~s$^{-1}$.
The velocity of the jet is $6C_{s0}\sim40$~km~s$^{-1}$, which is much larger than the reconnection outflow velocity.
These values are consistent with the observations.
The shock acceleration scenario can account for the difference between the velocity in the photosphere and the velocity of chromospheric jets.

\subsubsection{Ejections from Quiescent Prominences}
We found that the reconnection outflow speed increases through the interaction between one of the Petschek slow shocks and the steep density jump layer.
Recent observations have implied that magnetic reconnection takes place in quiescent prominences, where the quiescent prominences are large structure composed of relatively cool ($\sim 10^4$~K) plasma in the corona of the quiet-sun regions \citep{hil11}.
There are the density jump surfaces between the quiescent prominence and the corona.
Therefore, if the Petschek-type reconnection occurs in quiescent prominences, the reconnection outflow will be accelerated through the interaction.

\subsection{Justification of Anomalous Resistivity Model in Chromosphere}\label{anom}
We adopt an anomalous resistivity model not only in the corona but also in the chromosphere. 
The resistivity model is widely used to understand the rapid reconnection processes, namely the reconnection rate is close to unity, in a fully ionized and collision-less plasma like the corona \citep[e.g.][]{yok96}.
Using the resistivity model, many observational characteristics of explosive phenomena like solar flares and coronal jets have been successfully accounted for \citep[for review, see][]{shi11}.
If the anomalous resistivity model is adopted, the resistivity is spatially localized, which leads to the rapid Petschek-type reconnection.
Although there is no clear evidence that the Petschek slow shocks are indeed formed during reconnection processes, 
we can utilize the resistivity model to allow fast reconnection to occur.

\par
We should note that the chromosphere is fully collisional and partially ionized.
For this reason, the characteristics of the resistivity in the chromosphere can be different from that in the corona, 
which makes it difficult for us to predict the reconnection physics there.
Therefore before {\it Hinode} era  we had not known whether rapid reconnection processes would ubiquitously occur in the chromosphere.
Recently {\it Hinode} have observed many rapid events including the chromospheric anemone jets, which implies that rapid reconnection processes are taking place also in the chromosphere.

\par
It is known that the collision between electron, ion and neutral can play an important role in magnetic reconnection \citep{zwe97}. 
Neutrals mainly contribute to the gas pressure in a current sheet.
The collision between ions and neutrals leads to the diffusion of the neutrals from the current sheet.
This diffusion, i.e., the ambipolar diffusion, can case the current sheet to become thin.
When the thickness of the current sheet becomes comparable to a microscopic scale, magnetic reconnection will set in.
Recently, Isobe (in prep) showed that even with uniform resistivity, the localized (or non-uniform) neutral distribution in the current sheet induces the local thinning of the current sheet, leading to the Petschek-like reconnection.
This scenario may account for fast reconnection processes in the chromosphere.
Because there is a possibility that the Petschek-type reconnection takes place in the chromosphere,
we adopted the anomalous resistivity model to obtain it.
We believe that we can mimic reconnection processes in the partially ionized and fully collisional plasma using this resistivity model.

\subsection{Effect of Heat Conduction}\label{cnd_dis}
Generally speaking, the heat conduction weaken shocks because the heat energy liberated at the shock fronts can escape in front of the shock surfaces. 
We do not include the effect of the heat conduction in this paper. 
We consider whether or not the heat conduction significantly affects the simulation results.

\par
The timescale of the heat conduction can be estimated as
\begin{equation}
\tau_{cond}\sim 3\times 10^{-10}n_e l^2 T^{-5/2}\hspace{5mm}{\rm s},
\end{equation}
where $n_e$, $l$, $T$ are the electron density, typical length scale, and the temperature, respectively.
The heat conduction is important at a shock front because the temperature gradient is steep.
In our simulation, the minimum length scale is the smallest grid size.
By taking $l=8.5$~km (grid size), $n_e=10^{12}$~cm$^{-3}$ and $T=10^4$~K,
we estimate the heat conduction time $\tau_{cond}$ at $\sim2\times10^4$~s.
Considering that the propagation time of shocks in the chromosphere is less than 100~s in our simulation,
we can neglect the effect of the heat conduction on the shock propagation in the chromosphere.

\par
Heating due to the thermal conduction and cooling due to the radiative loss control the location of the transition region.
In slow shock acceleration scenario, the maximum height of a chromospheric jet is a function of the height of the transition region before a slow shock comes \citep{shi82}.
Therefore the maximum height of jets will be affected if we include the thermal conduction.
However, the acceleration scenario in which slow shocks lift the transition region will not be changed.

\par
The heating and cooling processes will affect the coronal EUV emission.
Therefore one should include the heat conduction as well as cooling terms into the energy equation if one tries to directly compare results of numerical simulations with observations like \citet{dep11} \citep[see e.g. ][]{heg09}.


\bigskip


We thank Dr. A. Hillier, Dr. S. Imada and Dr. Y. Katsukawa for fruitful discussion.
We also thank Dr. E. Priest for useful comment.
{\it Hinode} is a Japanese mission developed and launched by ISAS/JAXA, with NAOJ as domestic partner and NASA and STFC (UK) as international partners. It is operated by these agencies in co-operation with ESA and NSC (Norway).
This work was partially supported by the JSPS Core-to-Core Program 22001.
This work was supported by the Grand-in-Aid for the Global
COE Program ``The Next Generation of Physics, Spun from Universality \&
Emergence'' from the Ministry of Education, Culture, Sports, Science and
Technology (MEXT) of Japan.
HI is supported by the Grant-in-Aid for Young Scientists (B, 22740121).
The numerical calculations were carried out on SR16000 at  YITP in Kyoto University.
The authors thank the Yukawa Institute for Theoretical Physics at Kyoto University. Discussions during the YITP workshop YITP-W-12-08 on "Summer School on Astronomy \& Astrophysics 2012" were useful to complete this work.

\appendix
\section{1.5 Dimensional Study on Interaction between Slow Shock and Transition Region}

\begin{figure}
  \begin{center}
    \FigureFile(80mm,80mm){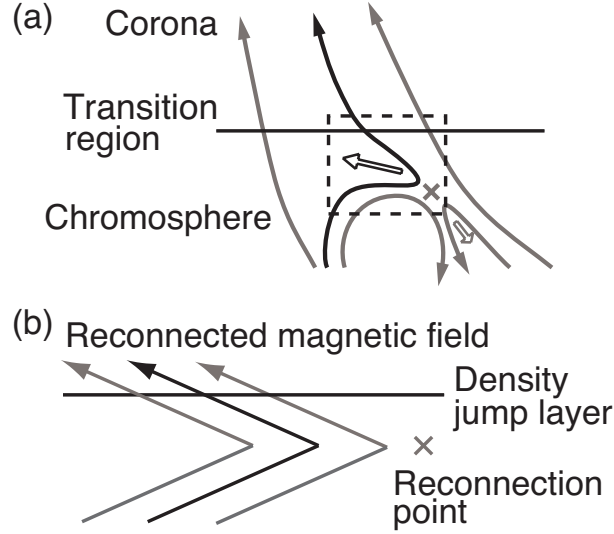}
  \end{center}
  \caption{Jet B: (a) the 2D situation discussed in Section~\ref{sec:upper}. (b) Simplification of the problem in which the density jump layer is located in the inflow region.}\label{2d1d}
\end{figure}

The process found in Section~\ref{sec:upper} can be considered as a magnetic reconnection problem with a density contact discontinuous layer (transition region) in the inflow region (see Figure~\ref{2d1d}).
We investigated the interaction process between the Petschek slow shock and the transition region.

\subsection{Basic Equations and Assumptions}
We simulated the time evolution of the reconnected magnetic field using 1.5D MHD simulations.
The gravity is neglected, therefore the results here are scale-free.
The basic equations are as follows:
\begin{eqnarray}
\frac{\partial \rho}{\partial t}+v_x\frac{\partial \rho}{\partial x}&=&-\rho \frac{\partial v_x}{\partial x}\\
\frac{\partial v_x}{\partial t} + v_x\frac{\partial v_x}{\partial x}&=&-\frac{1}{\rho}\frac{\partial p}{\partial x}-\frac{1}{4\pi \rho}B_y\frac{\partial B_y}{\partial x}\\
\frac{\partial v_y}{\partial t} + v_x\frac{\partial v_y}{\partial x} &=& \frac{1}{4\pi \rho}B_x \frac{\partial B_y}{\partial x}\\
\frac{\partial T}{\partial t} + v_x\frac{\partial T}{\partial x}&=&-(\gamma -1 )T\frac{\partial v_x}{\partial x}\\
p&=&\frac{k_B}{m}\rho T\\
\frac{\partial B_y}{\partial t}-c\frac{\partial E_z}{\partial x}&=&0\\
E_z &=& -\frac{v_x}{c} B_y+\frac{v_y}{c} B_x.
\end{eqnarray}

\subsection{Initial and Boundary Conditions}
\begin{figure}
  \begin{center}
    \FigureFile(60mm,80mm){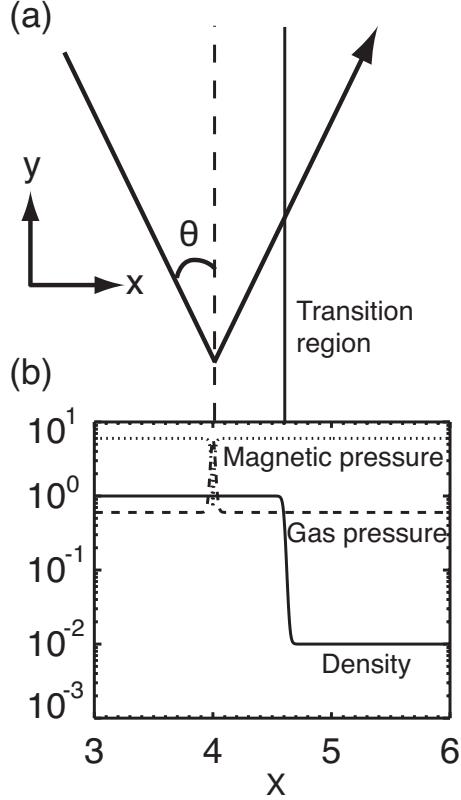}
  \end{center}
  \caption{The initial condition of the 1.5D simulation in the case that the ratio of the density in the chromosphere (high-density region) to the density in the corona (low-density region) is 100 and the plasma beta is 0.1.}\label{ic1.5}
\end{figure}

Figure~\ref{ic1.5} shows the initial condition of the simulation in the case that $\rho_{CH}/\rho_{CR}$ is 100, where $\rho_{CH}$ and $\rho_{CR}$ are the density in the chromosphere and the density in the corona, respectively.
The initial angle of the reconnected field line $\theta$, the plasma beta $\beta$ and the ratio of the density in the chromosphere and the density in the corona $\rho_{CH}/\rho_{CR}$ are parameters.
The parameters used are summarized in Table~\ref{tab:param}.
We investigated 48 cases.

\begin{longtable}{ll}
  \caption{Parameters}\label{tab:param}
  \hline              
  Parameter & Value \\ 
\endfirsthead
  \hline
  Parameter & Value  \\
\endhead
  \hline
\endfoot
  \hline
\endlastfoot
  \hline
$\rho_{CH}/\rho_{CR}$ & 10, 100, 1000\\
$\beta$ & 0.01, 0.1, 1, 10\\
$\theta$ [degrees] & 15, 20, 25, 30\\
\end{longtable}

\par
The initial distributions of the physical parameters are as follows:
\begin{eqnarray}
\rho &=& \rho_{CH} - (\rho_{CH} - \rho_{CR})\frac{1}{2}\left \{ \tanh{\left( \frac{x-x_d}{w} \right) }+1 \right \}\\
p &=&p_0 \left(1+\frac{1}{\beta}\right) - \frac{B_x^2+B_y^2}{8\pi}\\
B_x&=&B_0\sin{\theta}\\
B_y&=&B_0\cos{\theta}\tanh{\left( \frac{x-x_c}{w} \right)} \\
v_x&=&0\\
v_y&=&0,\\
\end{eqnarray}
where $B_0=\sqrt{8\pi p_0/\beta}$, $w=0.02$, $x_c=4$, $x_d=x_c+0.6$, $\rho_{CH}=1$ and $p_0=1/\gamma$.
The chromosphere and the corona correspond to the regions of $0\le x<4$ and $4<x$, respectively.
The velocity is normalized by the sound speed in the chromosphere.

\par
The size of the simulation box is $0\le x \le 240$.
The grid spacing is uniform with $\Delta x=0.0027$ in $0\le x \le 10$ and gradually increases in $10< x$.
The total grid number is 8200.
We use free boundary conditions.
The scheme used is a CIP-MOCCT scheme.

\subsection{Numerical Results}
\begin{figure}
  \begin{center}
    \FigureFile(160mm,80mm){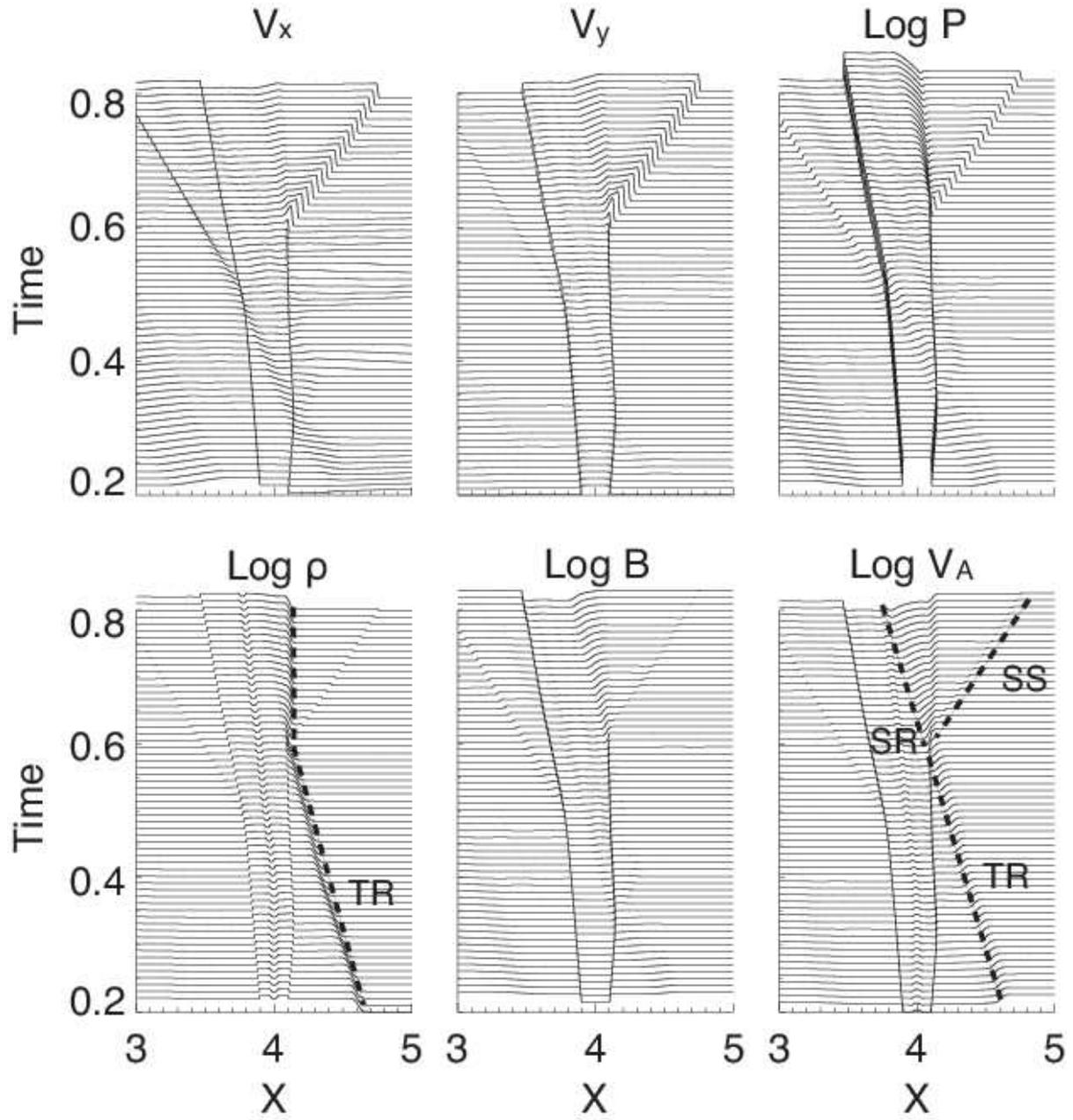}
  \end{center}
  \caption{Time evolution of the velocity in x-direction, the velocity in y-direction, the gas pressure, the density, the magnetic field strength and the Alfv\'en speed. TR, SS and SR denote the slow shock and the slow mode rarefaction wave, respectively.}\label{td6param}
\end{figure}

In Figure~\ref{td6param}, we show the time evolution of physical parameters for the case that $\rho_{CH}/\rho_{CR}=100$, $\beta=0.1$ and $\theta=20$~degrees.
We can discern two slow shocks are attached to the outflow region until $t\sim0.6$.
They corresponds to the Petchek slow shocks.
The transition region (TR) is advected by the inflow, which corresponds to a fast mode rarefaction wave, into the outflow region.
After one of the slow shocks crosses the transition region, 
the slow shock (SS) propagates in the positive x-direction.
Simultaneously, a slow mode rarefaction wave (SR), which corresponds to the reflection wave, is generated and propagates in the negative x-direction.

\par
\begin{figure}
  \begin{center}
    \FigureFile(60mm,80mm){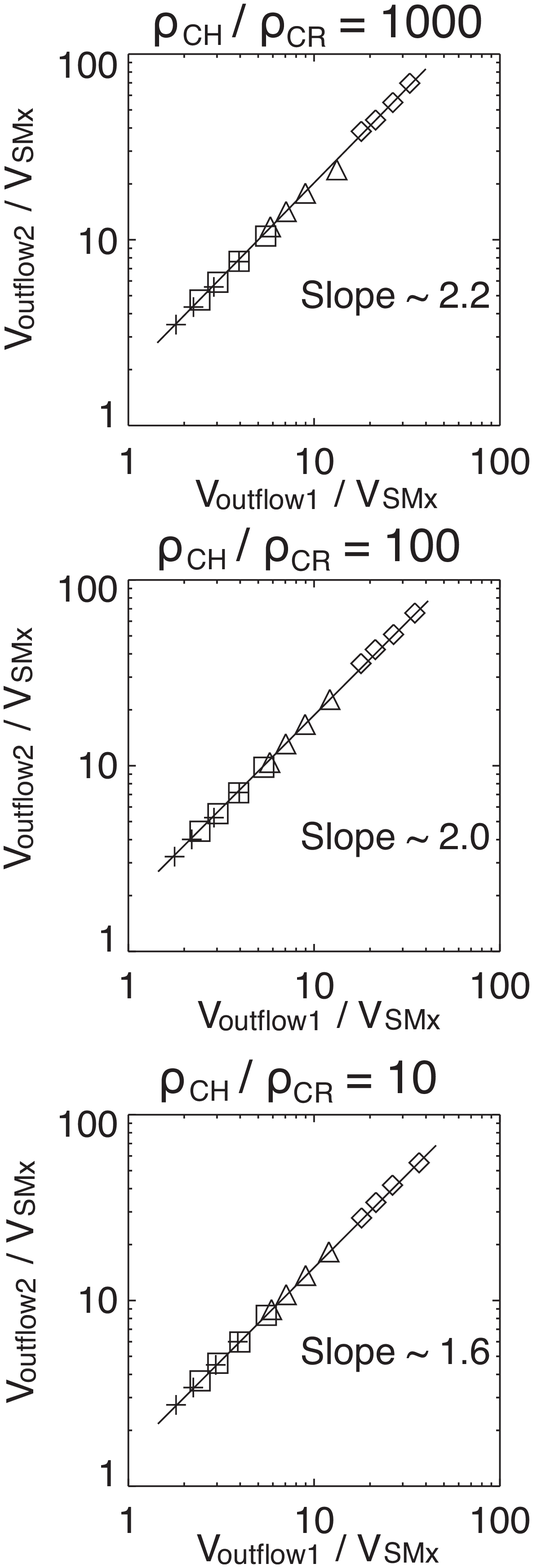}
  \end{center}
  \caption{Relation between the outflow velocity divided by the x-component of the unperturbed slow mode speed in the chromosphere before the interaction and that after the interaction. 
$\rho_{CH}$ and $\rho_{CR}$ are the density in the chromosphere and the density in the corona, respectively. 
The diamonds, triangles, squares and crosses are the results of the cases with $\beta$=0.01, 0.1, 1 and 10, respectively. 
The four points of each mark are the results of the cases with $\theta$=15, 20, 25 and 30 degrees, respectively.}\label{survey}
\end{figure}

From Figure~\ref{td6param}, we find that the outflow velocity increases after the interaction between the slow shock and the transition region.
We obtained a relation between the outflow velocity before the interaction $v_{\rm outflow1}$ and that after the interaction $v_{\rm outflow2}$: $v_{\rm outflow2} \sim 2v_{\rm outflow1}$ (see Figure~\ref{survey}).

\subsection{Interpretation of Numerical Results}
\begin{figure}
  \begin{center}
    \FigureFile(50mm,80mm){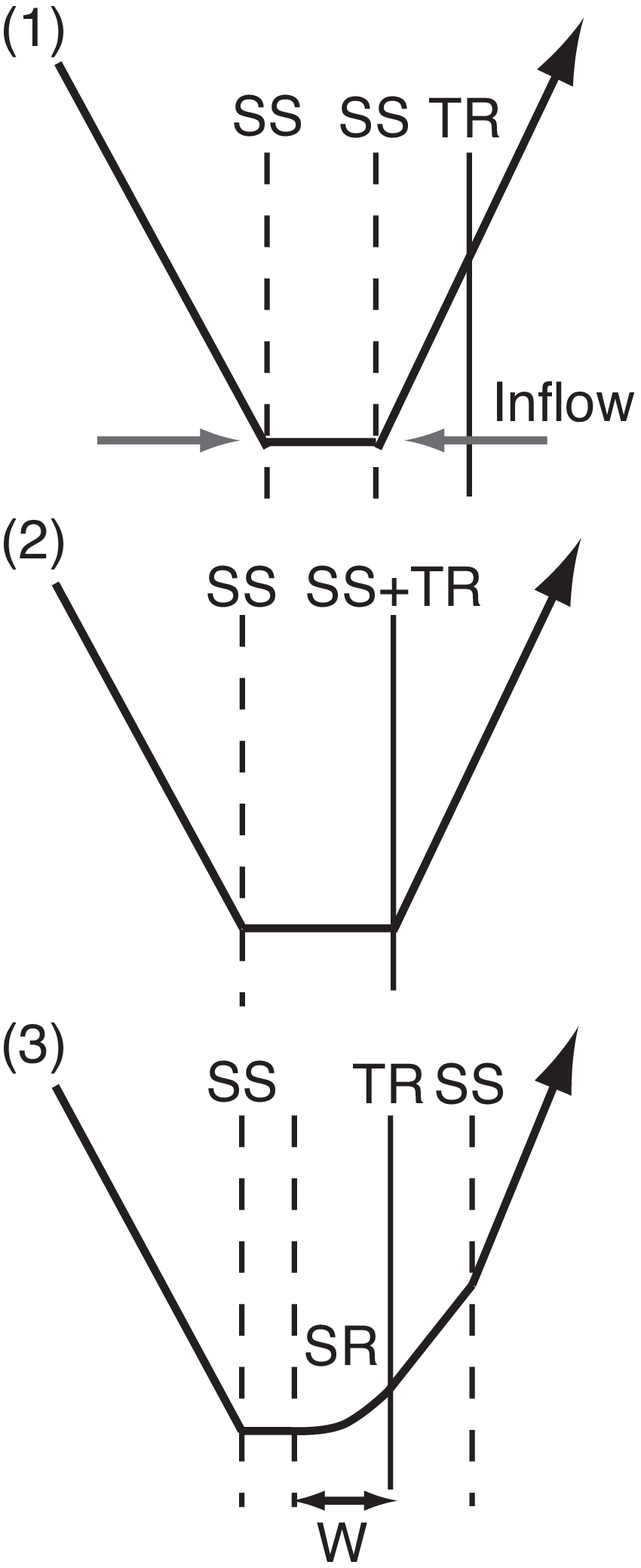}
  \end{center}
  \caption{Time evolution of the reconnected magnetic field.
 TR, SS and SR denote the transition region (density contact discontinuous layer), the slow shock and the slow mode rarefaction wave, respectively. }\label{1dponchi}
\end{figure}

Why does the outflow velocity increase through the interaction between the Petschek slow shock and the transition region?
We note that the slow mode rarefaction wave propagates into the outflow region after the interaction.
Behind the slow mode rarefaction wave, the magnetic field strength increases and the density decreases.
Thus the Alfv\'en speed increases.
The increase of the Alfv\'en speed can be found in our 2D simulation (see Figure~\ref{va2d}).
This leads to the rapid relaxation of the configuration of the magnetic field.
The re-configuration of the magnetic field causes the outflow plasma to be further accelerated.
The schematic diagram is shown in Figure~\ref{1dponchi}.

\par
Considering that the outflow plasma is further accelerated due to magnetic tension after the interaction,
we estimate the outflow velocity after the interaction as follows:

\begin{eqnarray}
\rho\frac{dv}{dt} &\sim& \frac{1}{4\pi}(\mbox{ \boldmath $B$}\cdot \nabla) \mbox{ \boldmath $B$}\\
\Delta v &\sim& \frac{B_x}{4\pi\rho}\frac{\partial B_y}{\partial x}\Delta t\\
&\sim& \frac{B_x}{4\pi \rho}\frac{B_y}{W}\frac{W}{v_{SMx}},
\end{eqnarray}
where $\Delta v$ is the increase of the outflow velocity through the interaction, $W$ is the length scale (see Figure~\ref{1dponchi}) and $v_{SMx}$ is the slow mode speed in the x-direction in the outflow region before the interaction.
In the outflow region where the magnetic field is in the x-direction,
\begin{equation}
v_{SMx} = \left \{
\begin{array}{l}
C_s \hspace{5mm} (\beta<2/\gamma)\\
V_{\it Ax} \hspace{5mm} (\beta>2/\gamma),\\
\end{array}
\right.
\end{equation}
where $V_{\it Ax}=B_x/\sqrt{4\pi\rho}$ is the Alfv\'en speed in x-direction.
For all the cases, the plasma beta in the outflow region is greater than $2/\gamma$. 
Thus $v_{SMx} = V_{\it Ax}$.
Therefore
\begin{eqnarray}
\Delta v &\sim& \frac{B_x}{4\pi \rho}\frac{B_y}{W}\frac{W}{v_{SMx}}\\
&=& \frac{B_x}{4\pi \rho}\frac{B_y}{W}\frac{W}{V_{\it Ax}}\\
&=& \frac{B_y}{\sqrt{4\pi\rho}}\\
&=&V_{\it Ay},
\end{eqnarray}
where $V_{\it Ay}=B_y/\sqrt{4\pi\rho}$ is the Alfv\'en speed in y-direction.
Before the interaction, the reconnection outflow velocity $v_{\rm outflow1}$ is $V_{\it Ay}$.
After the interaction, $v_{\rm outflow2}=V_{\it Ay}+\Delta v \sim 2V_{\it Ay}=2v_{\rm outflow1}$.
This relation shows no strong dependence on the plasma beta and the ratio of the density jump, which is consistent with the numerical results.



\begin{thebibliography}{}
\bibitem[Anan et al.(2010)]{ana10} Anan, T., Kitai, R., Kawate, T., et al.\ 2010, \pasj, 62, 871

\bibitem[Archontis \& Hood(2009)]{arc09} Archontis, V., \& Hood, A.~W.\ 2009, \aap, 508, 1469 

\bibitem[Archontis et al.(2010)]{arc10} Archontis, V., Tsinganos, K., \& Gontikakis, C.\ 2010, \aap, 512, L2

\bibitem[Bogdan et al.(2003)]{bog03} Bogdan, T.~J., Carlsson, M., Hansteen, V.~H., et al.\ 2003, \apj, 599, 626

\bibitem[Canfield et al.(1996)]{can96} Canfield, R.~C., Reardon, K.~P., Leka, K.~D., et al.\ 1996, \apj, 464, 1016

\bibitem[Carlsson \& Stein(1997)]{car97} Carlsson, M., \& Stein, R.~F.\ 1997, \apj, 481, 500

\bibitem[Chae et al.(1999)]{cha99} Chae, J., Qiu, J., Wang, H., \& Goode, P.~R.\ 1999, \apjl, 513, L75 

\bibitem[De Pontieu et al.(2004)]{dep04} De Pontieu, B., 
Erd{\'e}lyi, R., \& James, S.~P.\ 2004, \nat, 430, 536

\bibitem[De Pontieu et al.(2007)]{dep07} De Pontieu, B., 
Hansteen, V.~H., Rouppe van der Voort, L., van Noort, M., 
\& Carlsson, M.\ 2007, \apj, 655, 624 

\bibitem[De Pontieu et al.(2011)]{dep11} De Pontieu, B., 
McIntosh, S.~W., Carlsson, M., et al.\ 2011, Science, 331, 55

\bibitem[Ellerman(1917)]{ell17} Ellerman, F.\ 1917, \apj, 46, 298

\bibitem[Fang et al.(2006)]{fan06} Fang, C., Tang, Y.~H., Xu, Z., Ding, M.~D., \& Chen, P.~F.\ 2006, \apj, 643, 1325

\bibitem[Hansteen et al.(2006)]{han06} Hansteen, V.~H., De Pontieu, B., Rouppe van der Voort, L., van Noort, M., \& Carlsson, M.\ 2006, \apjl, 647, L73 

\bibitem[He et al.(2009)]{he09} He, J., Marsch, E., Tu, C., \& Tian, H.\ 2009, \apjl, 705, L217

\bibitem[Heggland et al.(2007)]{heg07} Heggland, L., De Pontieu, B., \& Hansteen, V. H. 2007, \apj, 666, 1277

\bibitem[Heggland et al.(2009)]{heg09} Heggland, L., De 
Pontieu, B., \& Hansteen, V.~H.\ 2009, \apj, 702, 1 

\bibitem[Hillier et al.(2011)]{hil11} Hillier, A., Isobe, H., \& Watanabe, H.\ 2011, \pasj, 63, L19

\bibitem[Isobe et al.(2006)]{iso06} Isobe, H., Miyagoshi, T., Shibata, K., \& Yokoyama, T.\ 2006, \pasj, 58, 423

\bibitem[Isobe et al.(2007)]{iso07} Isobe, H., Tripathi, D., \& Archontis, V.\ 2007, \apjl, 657, L53

\bibitem[Isobe et al.(2008)]{iso08} Isobe, H., Proctor, M.~R.~E., \& Weiss, N.~O.\ 2008, \apjl, 679, L57 

\bibitem[Jiang et al.(2011)]{jia11} Jiang, R.~L., Shibata, 
K., Isobe, H., \& Fang, C.\ 2011, \apjl, 726, L16 

\bibitem[Jiang et al.(2012)]{jia12} Jiang, R.-L., Fang, C., \& Chen, P.-F.\ 2012, \apj, 751, 152 

\bibitem[Kigure et al.(2010)]{kig10} Kigure, H., Takahashi, 
K., Shibata, K., Yokoyama, T., \& Nozawa, S.\ 2010, \pasj, 62, 993

\bibitem[Kitai(1983)]{kit83} Kitai, R.\ 1983, \solphys, 87, 135 

\bibitem[Kosugi et al.(2007)]{kos07} Kosugi, T., Matsuzaki, K., Sakao, T., et al.\ 2007, \solphys, 243, 3

\bibitem[Kudoh et al.(1999)]{kud99} Kudoh, T., Matsumoto, R., \& Shibata, K. 1999, Comput. Fluid Dyn. J., 8, 56

\bibitem[Leake et al.(2012)]{lea12} Leake, J.~E., Lukin, V.~S., Linton, M.~G., \& Meier, E.~T.\ 2012, \apj, 760, 109 

\bibitem[Mart{\'{\i}}nez-Sykora et al.(2009)]{mar09} 
Mart{\'{\i}}nez-Sykora, J., Hansteen, V., De Pontieu, B., 
\& Carlsson, M.\ 2009, \apj, 701, 1569

\bibitem[Mart{\'{\i}}nez-Sykora et al.(2011)]{mar11} Mart{\'{\i}}nez-Sykora, J., Hansteen, V., \& Moreno-Insertis, F.\ 2011, \apj, 736, 9

\bibitem[Matsumoto et al.(1993)]{mat93} Matsumoto, R., 
Tajima, T., Shibata, K., \& Kaisig, M.\ 1993, \apj, 414, 357

\bibitem[Matsumoto et al.(2008)]{mat08} Matsumoto, T., Kitai, R., Shibata, K., et al.\ 2008, \pasj, 60, 577 

\bibitem[Moreno-Insertis et al.(2008)]{mrn08} Moreno-Insertis, F., Galsgaard, K., \& Ugarte-Urra, I.\ 2008, \apjl, 673, L211

\bibitem[Morita et al.(2010)]{mor10} Morita, S., Shibata, K., Ueno, S., et al.\ 2010, \pasj, 62, 901

\bibitem[Murawski et al.(2011)]{mur11} Murawski, K., Srivastava, A.~K., \& Zaqarashvili, T.~V.\ 2011, \aap, 535, A58 

\bibitem[Nakamura et al.(2012)]{nak12} Nakamura, N., Shibata, K., \& Isobe, H.\ 2012, \apj, 761, 87 

\bibitem[Nishizuka et al.(2008)]{nis08} Nishizuka, N., Shimizu, M., Nakamura, T., et al.\ 2008, \apjl, 683, L83 

\bibitem[Nishizuka et al.(2011)]{nis11} Nishizuka, N., Nakamura, T., Kawate, T., Singh, K.~A.~P., \& Shibata, K.\ 2011, \apj, 731, 43 

\bibitem[Nozawa et al.(1992)]{noz92} Nozawa, S., Shibata, K., Matsumoto, R., et al.\ 1992, \apjs, 78, 267

\bibitem[{\^O}no et al.(1960)]{ono60} {\^O}no, Y., Sakashita, S., \& Yamazaki, H.\ 1960, Progress of Theoretical Physics, 24, 155

\bibitem[Osterbrock(1961)]{ost61} Osterbrock, D.~E.\ 1961, \apj, 134, 347

\bibitem[Parker(1966)]{parker66} Parker, E.~N.\ 1966, \apj, 145, 811

\bibitem[Pariat et al.(2004)]{pariat04} Pariat, E., Aulanier, G., Schmieder, B., et al.\ 2004, \apj, 614, 1099

\bibitem[Pariat et al.(2009)]{pariat09} Pariat, E., Antiochos, S.~K., \& DeVore, C.~R.\ 2009, \apj, 691, 61

\bibitem[Petschek(1964)]{pet64} Petschek, H.~E.\ 1964, NASA Special Publication, 50, 425

\bibitem[Roy(1973)]{roy73} Roy, J.~R.\ 1973, \solphys, 28, 95

\bibitem[Rust(1968)]{rust68} Rust, D.~M.\ 1968, Structure and Development of Solar Active Regions, 35, 77 

\bibitem[Saito et al.(2001)]{sai01} Saito, T., Kudoh, T., \& Shibata, K.\ 2001, \apj, 554, 1151

\bibitem[Schmieder et al.(1995)]{sch95} Schmieder, B., 
Shibata, K., van Driel-Gesztelyi, L., \& Freeland, S.\ 1995, \solphys, 156, 245 

\bibitem[Shibata \& Suematsu(1982)]{ss82} Shibata, K., \& Suematsu, Y. 1982, \solphys, 78, 333

\bibitem[Shibata et al.(1982)]{shi82} Shibata, K., Nishikawa, T., Kitai, R. \& Suematsu, Y. 1982, \solphys, 77, 121

\bibitem[Shibata et al.(1989)]{shi89} Shibata, K., Tajima, T., Steinolfson, R.~S., \& Matsumoto, R.\ 1989, \apj, 345, 584

\bibitem[Shibata et al.(1994)]{shi94} Shibata, K., Nitta, N., Strong, K.~T., et al.\ 1994, \apjl, 431, L51

\bibitem[Shibata et al.(2007)]{shi07} Shibata, K., Nakamura, T., Matsumoto, T., et al.\ 2007, Science, 318, 1591

\bibitem[Shibata \& Magara(2011)]{shi11} Shibata, K., \& Magara, T.\ 2011, Living Reviews in Solar Physics, 8, 6

\bibitem[Shimojo \& Shibata(2000)]{smj00} Shimojo, M., \& Shibata, K.\ 2000, \apj, 542, 1100

\bibitem[Singh et al.(2011)]{sin11} Singh, K.~A.~P., Shibata, K., Nishizuka, N., \& Isobe, H.\ 2011, Physics of Plasmas, 18, 111210

\bibitem[Singh et al.(2012)]{sin12} Singh, K.~A.~P., Isobe, H., Nishizuka, N., Nishida, K., \& Shibata, K.\ 2012, \apj, 759, 33

\bibitem[Sterling et al.(1993)]{ste93} Sterling, A. C., Shibata, K., \& Mariska, J. T. 1993, \apj, 407, 778

\bibitem[Sterling(2000)]{ste00} Sterling, A.~C.\ 2000, \solphys, 196, 79

\bibitem[Stone \& Edelman(1995)]{sto95} Stone, J.~M., \& Edelman, M.\ 1995, \apj, 454, 182

\bibitem[Suematsu et al.(1982)]{sue82} Suematsu, Y., Shibata, K., Neshikawa, T., \& Kitai, R.\ 1982, \solphys, 75, 99 

\bibitem[Takeuchi \& Shibata(2001)]{tak01} Takeuchi, A., \& Shibata, K.\ 2001, \apjl, 546, L73

\bibitem[Tsuneta et al.(2008)]{tsu08} Tsuneta, S., Ichimoto, K., Katsukawa, Y., et al.\ 2008, \solphys, 249, 167

\bibitem[Vernazza et al.(1981)]{valc} Vernazza, J.~E., Avrett, E.~H., \& Loeser, R.\ 1981, \apjs, 45, 635

\bibitem[Watanabe et al.(2008)]{wat08} Watanabe, H., Kitai, R., Okamoto, K., et al.\ 2008, \apj, 684, 736

\bibitem[Watanabe et al.(2011)]{wat11} Watanabe, H., Vissers, G., Kitai, R., Rouppe van der Voort, L., \& Rutten, R.~J.\ 2011, \apj, 736, 71 

\bibitem[Yokoyama \& Shibata(1996)]{yok96} Yokoyama, T., \& Shibata, K.\ 1996, \pasj, 48, 353

\bibitem[Yoshimura et al.(2003)]{yos03} Yoshimura, K., Kurokawa, H., Shimojo, M., \& Shine, R.\ 2003, \pasj, 55, 313 

\bibitem[Zweibel \& Brandenburg(1997)]{zwe97} Zweibel, E.~G., \& Brandenburg, A.\ 1997, \apj, 478, 563


\end{thebibliography}
\end{document}